\documentclass[english,notitlepage]{revtex4-1}
\usepackage[T1]{fontenc}
\usepackage[latin9]{inputenc}
\usepackage{amsmath}
\usepackage{amssymb}
\usepackage{graphicx}

\usepackage{dsfont}

\newcommand{\Extr}{\mbox{Extr}}
\newcommand{\Tr}{\mbox{Tr}}
\newcommand{\1}{\mathds{1}}

\usepackage{babel}

\begin{document}

\title{Optimal Learning with Excitatory and Inhibitory synapses}

\author{Alessandro Ingrosso}

\affiliation{Zuckerman Mind, Brain, Behavior Institute, Columbia University, New
York, NY, USA}

\begin{abstract}
Characterizing the relation between weight structure and input/output
statistics is fundamental for understanding the computational capabilities
of neural circuits. In this work, I study the problem of storing
associations between analog signals in the presence of correlations,
using methods from statistical mechanics. I characterize the typical
learning performance in terms of the power spectrum of random input
and output processes. I show that optimal synaptic weight configurations
reach a capacity of 0.5 for any fraction of excitatory to
inhibitory weights and have a peculiar synaptic distribution with
a finite fraction of silent synapses. I further
provide a link between typical learning performance and
principal components analysis in single cases. These results may
shed light on the synaptic profile of brain circuits, such as
cerebellar structures, that are thought to engage in processing
time-dependent signals and performing on-line prediction.
\end{abstract}
\maketitle

\section*{Introduction}
At the most basic level, neuronal circuits are characterized by the
subdivision into excitatory and inhibitory populations, a principle called Dale's law.
Even though the precise functional role of Dale\textquoteright s law
has not yet been understood, the importance of synaptic sign constraints
is pivotal in constructing biologically plausible models of synaptic
plasticity in the brain \cite{Song_training,NicolaClopathSupervised,Ingrosso_training,kimlearning,Deneve_spikebyspike}.
The properties of synaptic couplings strongly impact the
dynamics and response of neural circuits, thus playing a crucial role
in shaping their computational capabilities. It has been argued that
the statistics of synaptic weights in neural circuits could reflect
a principle of optimality for information storage, both at the
level of single-neuron weight distributions \cite{Brunel_optimal,Barbour_whatcanwelearn}
and inter-cell synaptic correlations \cite{Brunel_cortical} (e.g.
the overabundance of reciprocal connections). A number of theoretical
studies, stemming from the pioneering Gardner approach \cite{Gardner_spaceofinteractions},
have investigated the computational capabilities of stylized
classification and memorization tasks in both binary \cite{Clopath_storage,Chapeton_efficient,Zhang_associative,RubinBalanced}
and analog perceptrons \cite{Seung_learning,Clopath_optimal}, using
synthetic data. With some exceptions mentioned in the following,
these studies considered random uncorrelated inputs and outputs, a
usual approach in statistical learning theory. One interesting theoretical
prediction is that non-negativity constraints imply that a finite
fraction of synaptic weights are set to zero at critical capacity
\cite{Gutfreund_capacity,Brunel_optimal,Clopath_optimal}, a feature
which is consistent with experimental synaptic weight distributions
observed in some brain areas, e.g. input fibers to Purkinje cells
in the cerebellum.

The need to understand how the interaction between excitatory and inhibitory synapses
meditates plasticity and dynamic homeostasis \cite{Isaacson_inhibition,Field_HeterosynapticPlasticity}
calls for the study of heterogeneous multi-population feed-forward
and recurrent models. A plethora of mechanisms for excitatory-inhibitory
(E-I) balance of input currents onto a neuron have been proposed
\cite{Hennequin_review,ahmadian_dynamical}.
At the computational level, it has recently been shown that a peculiar
scaling of excitation and inhibition with network size, originally
introduced to account for the high variability of neural firing activity
\cite{VanVreeswijkChaosScience,VanVreeswijkChaoticBalancedStateNeuralComputation,RenartAsynchronous,KadmonSompolinsky,HarishHanselAsynchronous,BrunelDynamicsSparsely,TsodyksStateSwitching},
carries the computational advantage of noise robustness and stability
of memory states in associative memory networks \cite{RubinBalanced}.

Analyzing training and generalization performance
in feed-forward and recurrent networks as a function of statistical
and geometrical structure of a task remains an open problem both in
computational neuroscience and statistical learning theory \cite{goldt_modelling,Chung_perceptualmanifold,Cohen_Separability}.
This calls for statistical models of the low-dimensional structure of data that
are at the same time expressive and amenable to mathematical analyses.
A few classical studies investigated the effect of ``semantic'' (among input patterns)
and spatial (among neural units) correlations in random classification
and memory retrieval \cite{Monasson_properties,Tarkowski_optimal,Monasson_correlatedpatterns}.
The latter are important in the construction of associative
memory networks for place cell formation in the hippocampal complex
\cite{Battista_capacity}.

For reason of mathematical tractability, the vast majority of
analytical studies in binary and analog perceptron models focused
on the case where both inputs and outputs are independent and identically
distributed. In this work, I relax this assumption and study optimal
learning of input/output associations with real-world statistics
with a linear perceptron having heterogeneous synaptic weights.
I introduce a mean-field theory of an analog perceptron in the presence
of weight regularization with sign-constraints, considering
two different statistical models for input and output correlations.
I derive its critical capacity in a random association task and study
the statistical properties of the optimal synaptic weight vector across
a diverse range of parameters.

This work is organized as follows. In the first section, I introduce
the framework and provide the general definitions for the problem.
I first consider a model of temporal (or, equivalently, \textquotedblleft semantic\textquotedblright)
correlations across inputs and output patterns, assuming statistical
independence across neurons. I show that optimal solutions are insensitive
to the fraction of E and I weights, as long as the external bias is
learned. I derive the weight distribution and
show that it is characterized by a finite fraction
of zero weights also in the general case of E-I constraints and correlated signals.
The assumption of independence is subsequently relaxed
and I build on self-averaging adaptive TAP formalism
to provide a theory that depends on the spectrum of the sample covariance
matrix and the dimensionality of the output signal along the principal
components of the input. The implications of these results are
discussed in the final section.


\section*{Results}

\subsection*{Mean-field theory with correlations}
\begin{figure}
\centering
\includegraphics[scale=1]{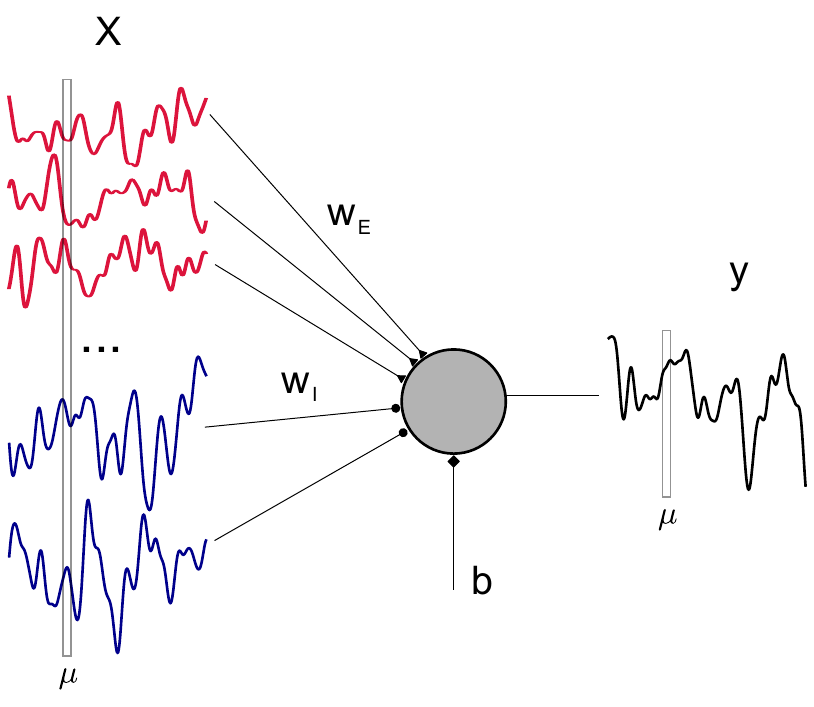}
\caption{{\bf Schematic of the learning problem.}
A linear perceptron
receives $N$ correlated signals (input rates of pre-synaptic neurons) $x_{i\mu}$ and maps them
to the output $y_{\mu}$ through $N_E=f_{E}N$ excitatory and $N_I=\left(1-f_{E}\right)N$
plastic inhibitory weights $w_{i}$, plus an additional bias current
$b$.}
\label{fig:fig1}
\end{figure}

Consider the problem of linearly mapping a set of time-correlated
signals $x_{i\mu}$, with $i\in1,...,N$ and $\mu=1,...,M$ from $N_E=f_{E}N$
excitatory (E) and $N_I=\left(1-f_{E}\right)N$ inhibitory (I) neurons, onto an output
signal $y_\mu$ using a synaptic vector $\boldsymbol{w}$, in the presence
of a learnable constant bias current $b$ (Fig \ref{fig:fig1}).
To account for different statistical properties of E and
I input rates, we write the elements of the input matrix as $\left(X\right)_{i\mu}\equiv x_{i\mu}=\bar{x}_{i}+\sigma_{i}\xi_{i\mu}$
with $\bar{x}_{i}=\bar{x}_{E}$ for $i\leq f_{E}N$ and $\bar{x}_{i}=\bar{x}_{I}$
for $i>f_{E}N$ and the same for $\sigma_{i}$. At this stage, the quantities $\xi_{i\mu}$
have unit variance and are uncorrelated across neurons: $\left\langle \xi_{i\mu}\xi_{i\nu}\right\rangle =\delta_{ij}C_{\mu\nu}$.
The output signal has average $\left\langle y_{\mu}\right\rangle =\bar{y}$
and variance $\left\langle \left(y_{\mu}-\bar{y}\right)^{2}\right\rangle =\sigma_{y}^{2}$.
We initially consider output signals $y_{\mu}$ with the same
temporal correlations as the input, namely $\left\langle \delta y_{\mu}\delta y_{\nu}\right\rangle =C_{\mu\nu}$,
where $y_{\mu}=\bar{y}+\sigma_y\delta y_{\mu}$. For a given input-output
set, we are faced with the problem of minimizing the following regression loss (energy) function:
\begin{equation}
E\left(\boldsymbol{w};\gamma,x,y\right)=\frac{1}{2}\sum_{\mu=1}^{M}\left(\sum_{i=1}^{N}w_{i}x_{i\mu}+b-y_{\mu}\right)^{2}+\frac{N\gamma}{2}\sum_{i=1}^N w_{i}^{2}
\label{eq:loss_function}
\end{equation}
with $w_{i}>0$ for $i\leq f_{E}N$, $w_{i}<0$ otherwise. The typical
vector $\boldsymbol{w}$ that solves this sign-constrained least square problem
has a squared norm $\sum_{i=1}^{N}w_{i}^{2}=\mathcal{O}\left(1\right)$,
hence the scaling of the regularization term $N\gamma$. In order
to consider a well defined $N\to\infty$ limit for $E$ and the spectrum
of the matrix $C$, we take $M=\alpha N$, with $\alpha$ called the load,
as is costumary in mean-field analysis of perceptron problems \cite{Gardner_spaceofinteractions}.
Numerical experiments show that the optimal bias current is of order
$\sqrt{N}$, as can be derived in the special case of i.i.d input/output
and non-negative synaptic weights $w_{i}$ \cite{Clopath_optimal}.
Optimizing with respect to the bias $b=I\sqrt{N}$ naturally yields
solutions $\boldsymbol{w}$ for which
\begin{equation}
N_E\bar{w}_{E}\bar{x}_{E}+N_I\bar{w}_{I}\bar{x}_{I}+b=\bar{y}
\label{eq:balance}
\end{equation}
where we call $\bar{w}_{c}=\frac{1}{N_c}\sum_{i\in c}w_{i}=\mathcal{O}\left(1/\sqrt{N}\right)$
the average excitatory and inhibitory weight, with $c\in\left\{E,I\right\}$.
We call this property \emph{balance}, in that the same scaling is used in balanced state
theory of neural circuits \cite{VanVreeswijkChaosScience,VanVreeswijkChaoticBalancedStateNeuralComputation,KadmonSompolinsky}.

In order to derive a mean-field description for the typical properties
of the learned synaptic vector $\boldsymbol{w}$, we employ a statistical
mechanics framework in which the minimizer of $E$ is evaluated after
averaging across all possible realizations of the input matrix $X$ and
output $y$. To do so, we compute the free energy density
\begin{equation}
f=-\frac{1}{\beta N}\left<\log \int {d\mu\left(\boldsymbol{w}\right)}e^{-\beta E}\right>_{x,y}
\label{eq:free-energy}
\end{equation}
where $d\mu\left(\boldsymbol{w}\right)$ is the measure implementing the sign-constraints
over the synapic weight vector $\boldsymbol{w}$. The brackets in Eq~(\ref{eq:free-energy})
stand for the quenched average over all the quantities
$x_{i\mu}$ and $y_\mu$, and the inverse temperature $\beta$ will allow
us to select weight configurations $\boldsymbol{w}$ which minimize the energy $E$.
The free energy $f$ acts as a generating function from which all the statistical quantities
of interest can be calculated by appropriate differentiation and taking the $\beta\to\infty$ limit.
In particular, we will be interested in the average loss $\left<E\right>$ and
the error $\epsilon=\frac{1}{2}\left<\left|X^T \boldsymbol{w} + b - \boldsymbol{y}\right|^2\right>$,
which corresponds to the average value of the first term in Eq~(\ref{eq:loss_function}).
The average in Eq~(\ref{eq:free-energy}) can be computed in the $N\to\infty$ limit
with the help of the replica method, an analytical continuation technique
that entails the introduction of a number $n$ of \emph{formal} replicas
of the vector $\boldsymbol{w}$. A general expression for $f$ can be
obtained in the large $N$ limit using the saddle point method.
The crucial quantity in our derivation
is the (replicated) cumulant generating function $Z_{\xi,\delta y}$ for
the (mean-removed) input $x$ and output $y$, which can be easily
expressed as a function of the eigenvalues $\lambda_{\mu}$, $\mu=1,...,\alpha N$
of the covariance matrix $C$, plus a set of order parameters to be
evaluated self-consistently (Methods).

\subsection*{Critical capacity}
The existence of weight vectors $\boldsymbol{w}$'s with a certain value of the
regression loss $E$ in the error regime ($E>0$) is described by the order parameter $\Delta\tilde{q}_w$.
For finite $\beta$, the quantity $\Delta q_w = \beta \Delta\tilde{q}_w$ represents the
variance of the synaptic weights across different
solutions. In the asymptotic limit $\beta\to\infty$ of Eq~(\ref{eq:free-energy}),
a simple saddle point equation for $\Delta\tilde{q}_w$ can be derived when $b$ is
chosen to minimize Eq~(\ref{eq:loss_function}):
\begin{equation}
\alpha\Delta\tilde{q}_w\left\langle \frac{\lambda}{1+\Delta\tilde{q}_w\lambda}\right\rangle_{\rho\left(\lambda\right)}=\frac{1}{2}-\gamma\Delta\tilde{q}_w
\label{eq:delta_q_tilde}
\end{equation}
where $\rho\left(\lambda\right)$ is the distribution of eigenvalues of $C$.

\begin{figure}
\centering
\includegraphics[scale=1]{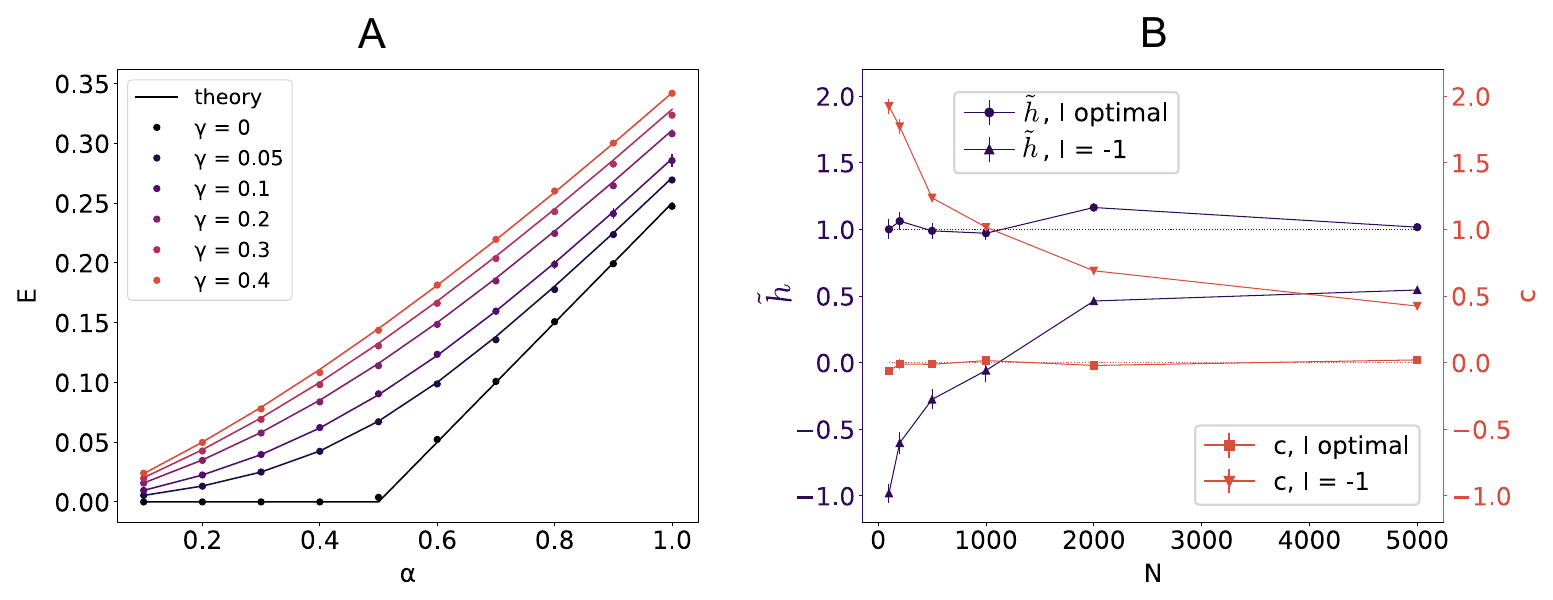}
\caption{{\bf Critical capacity and weight balance.}
A: Average loss $\left<E\right>$ for a linear perceptron with $f_{E}=0.8$ positive synaptic weights
in the case of i.i.d. input $X$ and output $y$ for increasing values
of the regularization $\gamma$. Parameters: $N=1000$, $\bar{x}_{E}=\bar{x}_{I}=\sigma_{E}=\sigma_{I}=\bar{y}=\sigma_{y}=1$.
Each point is an average across $50$ samples. Full lines show the
theoretical results. B: Mean-field component $\tilde{h}$ (left axis, purple)
and weight-input correlation $c$ (right axis, red) for increasing dimension $N$
in the case where the bias current $b=I\sqrt{N}$
is either learned ($I$ optimal) or fixed at the outset ($I=-1$) for $f_{E}=1$, $\gamma=0.1$, $\alpha=0.8$.
Inputs $X$ and output $y$ are time-correlated with un-normalized Gaussian covariance $C$, $\tau=10$ (see text).
The remaining parameters are as in A. The asymptotic value $\tilde{h}=\bar{y}=1$ is highlighted by the purple dotted line,
the value $c=0$ by the red dotted line as guide for the eye.}
\label{fig:fig2}
\end{figure}

In the absence of weight regularization ($\gamma=0$), we define
the critical capacity $\alpha_{c}$ as the maximal load $\alpha=M/N$
for which the patterns $\boldsymbol{x}_{\mu}$ can be correctly mapped to their
outputs $y_{\mu}$ with zero error. When the synaptic weights are not sign-constrained,
the critical capacity is obviously $\alpha_{c}=1$, since the matrix
$X$ is typically full rank. In the sign-constrained case, $\alpha_{c}$
is found to be the minimal value of $\alpha$ such that Eq~(\ref{eq:delta_q_tilde})
is satisfied for $0<\Delta\tilde{q}_w<\infty$.
Noting that the left-hand side in Eq~(\ref{eq:delta_q_tilde}) is a non-decreasing function of $\Delta\tilde{q}_w$
with an asymptote in $\alpha$, the order parameter $\Delta\tilde{q}_w$
goes to $\infty$ as the critical capacity is approached from the right.
We thus find for $\gamma=0$ the surpisingly simple result:
\begin{equation}
\alpha_c = 0.5
\label{eq:critical_capacity}
\end{equation}
As shown in Fig \ref{fig:fig2}A in the case of i.i.d. $x$ and $y$,
the loss has a sharp increase at $\alpha=0.5$. This holds
irrespectively of the structure of the covariance matrix $C$ and
the ratio of excitatory weights $f_{E}$. In Fig \ref{fig:fig2}A, we also show the average
minimal loss $E$ for increasing values of the regularization parameter
$\gamma$.

For a generic value of the bias current $b$, there are strong deviations from the condition in Eq~(\ref{eq:balance}).
In Fig \ref{fig:fig2}B, we compare the value of the average output $\bar{y}$
with $\tilde{h}\equiv \sum_{c\in\left\{E,I\right\}}N_c\bar{w}_{c}\bar{x}_{c}+b$,
and also plot the residual term $c=\frac{1}{NM}\sum_{i\mu}\delta w_{i}x_{i\mu}$,
where we decomposed the weight vector components as $w_{i}=\bar{w}_{c}+\delta w_{i}$
for $c\in\left\{ E,I\right\} $. The quantity $c$ measures weight-rate correlations
which are responsible for the cancelation of the $\mathcal{O}\left(\sqrt{N}\right)$
bias.

The deviation from Eq~(\ref{eq:balance}),
shown here for a rapidly decaying covariance of the form $C_{\mu\nu}=e^{-\frac{\left|\mu-\nu\right|}{2\tau^{2}}}$,
has been previously described in the context of a target-based learning
algorithm used to build E-I-separated rate and spiking models of neural circuits capable
of solving input/output tasks~\cite{Ingrosso_training}.
In this approach, a randomly initialized recurrent networks $n_T$ is
driven by a low dimensional signal $z$. Its currents
are then used as targets to train the synaptic couplings of a second (rate or spiking)
network $n_S$, in such a way that the desired output $z$ can later be linearly decoded
from the self-sustained activity of $n_S$.
Each neuron of $n_S$ has to independently learn an input/output mapping
from firing rates $x$ to currents $y$, using an on-line sign-constrained least square
method. In  the presence of an L2 regularization and a constant $b\propto\sqrt{N}$ external
current, the on-line learning method typically converges onto a solution
for the recurrent synaptic weights for which Eq~(\ref{eq:balance}) does not hold.
As also shown in~\cite{Ingrosso_training}, in the peculiar case of a self-sustained periodic dynamics 
(in which case off-diagonal terms of the covariance matrix $C_{\mu\nu}$ do
not vanish for large $\mu$ or $\nu$) the two contributions $\tilde{h}$
and $c$ scale approximately like $\sqrt{N}$ and cancel each other
to produce an $\mathcal{O}\left(1\right)$ total average output $\bar{y}=\tilde{h}+c$.

\subsection*{Power spectrum and synaptic distribution}
\begin{figure}
\centering
\includegraphics[scale=1]{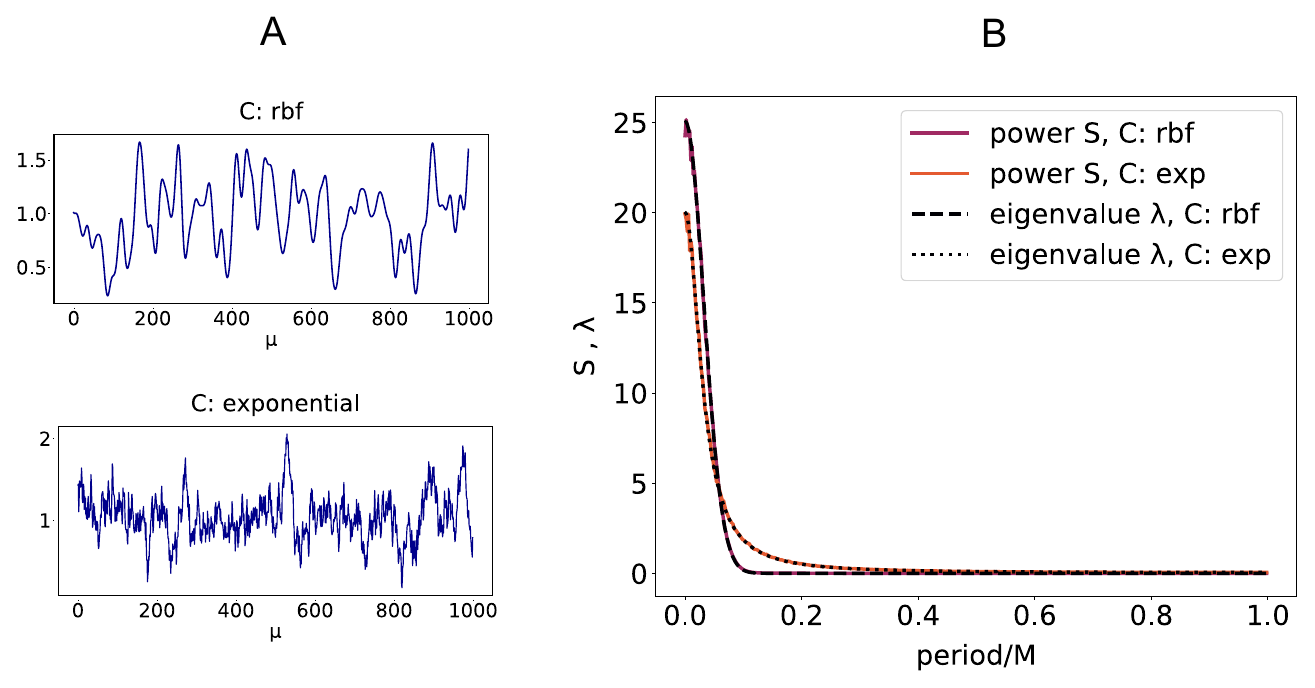}
\caption{{\bf Eigenvalues of $C$ and Fourier spectrum.}
A: Examples of excitatory input signals $x_{i\mu}$ ($i \in E$)
with two different covariance matrices $C$.
Top: rbf covariance, $\tau=10$. Bottom: exponential covariance $C_{\mu\nu}=e^{-\frac{\left|\mu-\nu\right|}{\tau}}$, $\tau=10$.
Parameters: $\bar{x}_E=1$, $\sigma_E=0.3$.
B: Theoretical eigenvalue spectrum of $C$ with $\tau=10$ versus average power spectrum
for positive wave numbers across $N=2000$ independent processes with $M=1000$ time steps.}
\label{fig:fig_eigs}
\end{figure}

\begin{figure}
\centering
\includegraphics[scale=1]{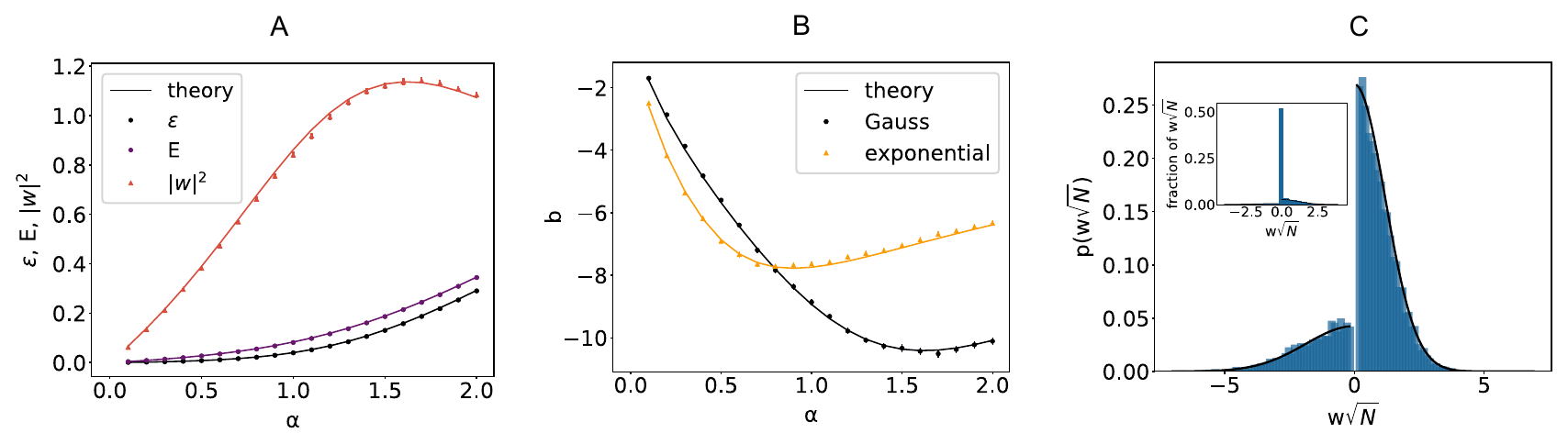}
\caption{{\bf Learning temporally structured signals.}
A: Minimal loss $E$, error $\epsilon$
and norm of the weight vector $\boldsymbol{w}$ as a function of the load $\alpha$
for a linear perceptron trained on a time-correlated signal. Covariance matrix
$C$ is of rbf type with $\tau=2$. Parameters: $N=1000$,
$f_{E}=0.8$, $\gamma=0.1$, $\bar{x}_{E}=\bar{x}_{I}=\sigma_{E}=\sigma_{I}=\bar{y}=\sigma_{y}=1$.
B: Optimal bias $b$ for the two sets of signals with rbf
(black curve) and exponential (yellow curve) covariance $C$, with
$\tau=2$. Theoretical curves show the value $I\sqrt{N}+\bar{y}$, where $I$
has been computed form the saddle point equations (Methods).
Parameters as in A. Each point in A and B is an average
across $50$ samples.
C: Probability density of non-zero synaptic weights $w_i\sqrt{N}$ of a linear perceptron with $N=1000$, a
fraction $f_{E}=0.8$ of excitatory weights, trained on $M=600$ exponentially
correlated input $x$ and output $y$.
The $\delta$ function in zero is omitted for better visualization.
Parameters: $\tau=10$, $\gamma=0.1$,
$\bar{x}_{E}=\bar{x}_{I}=1$, $\sigma_{I}=2\sigma_{E}=0.4$.
The histogram is an average across 50 realizations of input/output
signals. Inset: full histogram of synaptic weights $w_i\sqrt{N}$.}
\label{fig:fig3}
\end{figure}

The theory developed thus far applies to a generic covariance matrix $C$.
To connect the spectral properties of $C$ with the signal dynamics,
we further assume the $x_{i\mu}$ to be $N$ independent stationary
discrete-time processes. In this case, $C_{\mu\nu}=C\left(\mu-\nu\right)$
is a matrix of Toeplitz type~\cite{Gray_toeplitz}, leading to the following expression
for the average minimal energy in the $N\to\infty$ limit: 
\[
\left<E\right>=\frac{\sigma^2_y}{2\pi}\int_0^\pi d\phi\frac{\lambda\left(\phi\right)}{1+\Delta\tilde{q}_w\lambda\left(\phi\right)}
\]
with $\Delta\tilde{q}_w$ given by Eq~(\ref{eq:delta_q_tilde}).
The function $\lambda\left(\phi\right)$ can be computed exactly in some cases
(Methods) and corresponds to the average power spectrum of the
$x$ and $y$ stochastic processes. Fig~\ref{fig:fig_eigs} shows
two representative input signals with Gaussian and exponential covariance
matrix $C$ (Fig~\ref{fig:fig_eigs}A) and a comparison between
the average power spectrum of the input
and the analytical results for the eigenvalue spectrum of the matrix $C$ (Fig~\ref{fig:fig_eigs}B).
From now on, we use the terms Gaussian or rfb (radial basis function) indistinguishably to denote the un-normalized
Gaussian function $C_{\mu\nu}=e^{-\frac{\left(\mu-\nu\right)^{2}}{2\tau^{2}}}$.
As shown in Fig~\ref{fig:fig3}A in the case of input $x$ and output $y$
with rbf covariance, the squared norm of the optimal synaptic
vector $\boldsymbol{w}$ (red curve) is in general a non-monotonic function of $\alpha$,
its maximum being attained at bigger values of $\alpha$ as the time constant $\tau$ increases.
We also show the minimal energy $E$ and the mean error $\epsilon$ for $\gamma=0.1$.
The curves in Fig~\ref{fig:fig3}A are the same for any ratio $f_{E}$:
the use of an optimal bias current $b$ cancels any asymmetry between E and I populations.
For a finite $\gamma$, the average minimal energy $E$ for a given
$f_{E}$ decreases as either $\sigma_{E}$ or $\sigma_{I}$ increase.
For a given set of parameters $f_{E}$ and $\gamma$, the optimal
bias $b$ will in general depend on the load $\alpha$ and
the structure of the covariance matrix $C$, as shown in Fig~\ref{fig:fig3}B.
The probability distribution of the weights $w_i$ can
easily be derived using the same analytical machinery
employed for the calculation of the free-energy $f$ (Methods).
For a fixed bias $b$, the probability density of the synaptic weights
is composed of two truncated Gaussian distributions with zero mean for the E and I
components, plus a finite fraction $p_{0}$ of zero weights,
given by 
\begin{equation}
p_{0}\left(B\right)=f_{E}H\left(-\eta_E B\right)+\left(1-f_{E}\right)H\left(\eta_I B\right)
\end{equation}
where $H\left(x\right)=\int_{x}^{\infty}dz \frac{e^{-\frac{z^2}{2}}}{\sqrt{2\pi}}$, $B$ is an order
parameter that must to be computed from the saddle point equations, and
$\eta_{c}=\frac{\bar{x}_{c}}{\sigma_{c}}$, with $c\in\left\{ E,I\right\}$.
Interestingly, the optimal bias $b$ yields the simple
results $B=0$, which greatly simplifies the saddle point equations,
and implies that half of the synapses are zero, irrespectively of
$f_{E}$ and the properties of the covariance matrix $C$. We show
in Fig~\ref{fig:fig3}D the shape of the optimal weight distribution
for a linear perceptron with $80\%$ excitatory synapses, trained on exponentially
correlated $x$ and $y$ and with a ratio $\sigma_{I}/\sigma_{E}=2$.
It is interesting to note that, in the presence of an optimal external current,
both the means of the Gaussian components and the fraction of
silent synapses do not depend on the specific properties
of input and output signals.

\begin{figure}
\centering
\includegraphics[scale=1]{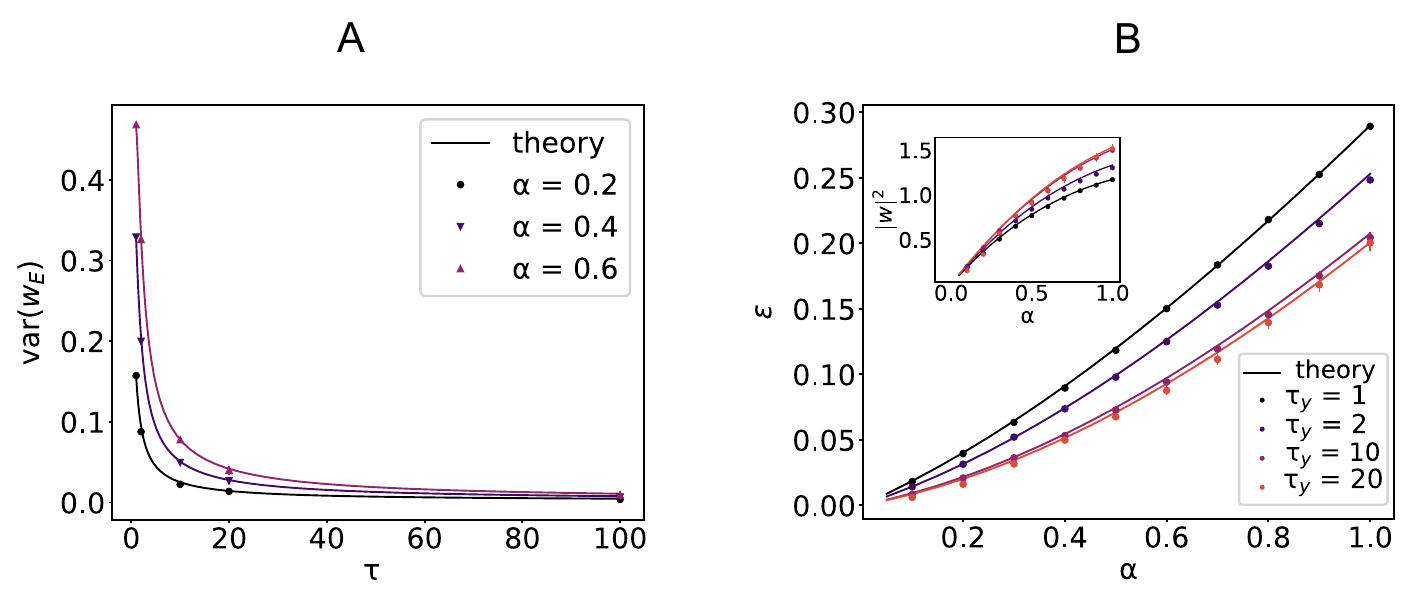}
\caption{{\bf Input/output time constants and learning performance.}
A: Variance of synaptic weights ($f_{E}=1$) for
a linear perceptron of dimension $N=1000$ trained on rbf-correlated signals
with increasing time constant $\tau$ for three different values of
the load $\alpha$. Parameters: $\gamma=0.1$, $\bar{x}_{E}=\bar{x}_{I}=\sigma_{E}=\sigma_{I}=\bar{y}=\sigma_{y}=1$.
B: Average error $\epsilon$ in the case where input and output signals
have two different covariance matrices, for increasing time constant
$\tau_{y}$ of the output signal $y$. Parameters: $N=1000$, $f_{E}=0.8$,
$\gamma=0.1$, $\bar{x}_{E}=\bar{x}_{I}=\bar{y}=\sigma_{y}=1$, $\sigma_{I}=2\sigma_{E}=0.6$,
$C^{x}$ rbf with $\tau_{x}=1$, $C^{y}$ rbf with various values of
$\tau_{y}$. Inset: norm of the weight
vector $\boldsymbol{w}$. Full lines show analytical results. Points are averages
across $50$ samples.}
\label{fig:fig4}
\end{figure}

The dynamic properties of input/output mappings affect the shape of
the weight distribution in a computable manner.
As an example, in a linear perceptron with non-negative synapses, the
explicit dependence of the variance of the weights on the input and output auto-correlation time
constant is shown in Fig~\ref{fig:fig4}A for various loads
$\alpha$. Previous work considered an analog perceptron with purely
excitatory weights as a model for the graded rate response of Purkinje
cells in the cerebellum~\cite{Clopath_optimal}.
In the presence of heterogeneity of synaptic properties across cells,
a larger variance in their synaptic distribution
is expected to be correlated with high frequency temporal fluctuations in input currents.
Analogously, the auto-correlation of the typical signals being processed 
sets the value of the constant external current that a neuron
must receive in order to optimize its capacity.

When the input and output have different covariance matrices $C^{x}\neq C^{y}$,
a joint diagonalization is not possible in general (Methods).
We can nevertheless write an expression (Eq~(\ref{eq:app_energetic_part_cov_xy}))
that holds when input and output patterns are defined on a ring (with periodic boundary conditions)
and use it as an approximation for the general case.
Fig~\ref{fig:fig4}B shows good agreement between numerical experiment and theoretical
predictions for the error $\epsilon$ and the squared norm of the synaptic
weight vector $\boldsymbol{w}$, when input and output processes have two different
time-constants $\tau_{x}$ and $\tau_{y}$.

\subsection*{Sample covariance and dimensionality}
In the discussion thus far, we assumed independence across the ``spatial''
index $i$ in the input. It is often the case for input signals to
be confined to a manifold of dimension smaller than $N$, a feature
that can be described by various dimensionality measures,
some of which rely on principal component analysis~\cite{abbott_interactions,LitwinKumar_optimal}.
In order to relax the independence assumption, we build on a framework
originally introduced in the theory of spin glasses with orthogonal
couplings \cite{Marinari_replicafield,Parisi_orthogonal,Cherrier_interactionmatrix}
and further developed in the context of adaptive TAP equations \cite{Opper_tractable,Opper_adaptive,Opper_expectation}.
Following previous work in the context of information theory of linear
vector channels and binary perceptrons \cite{Takeda_cdma,Kabashima_unified,Shinzato_learning,Shinzato_Revisited},
we employ an expression for an ensemble of rectangular random matrices.

Let us write the input matrix $\left(X\right)_{i\mu}=\bar{x}_{i}+\sigma_{i}\xi_{i\mu}$,
with $\xi=USV^{T}$, $S$ being the matrix of singular values. 
To analyze the properties of the typical case, we start from
a generic singular value distribution $S$ and consider i.i.d. output
$y_\mu$. In calculating the cumulant generating function $Z_{\xi,\delta y}$,
we perform a homogeneous average across the left and right principal
components $U$ and $V$. Calling $\rho_{\xi\xi^{T}}\left(\lambda\right)$ 
the eigenvalue distribution of the sample covariance matrix $\xi\xi^{T}$,
we can express $Z_{\xi,\delta y}$ in terms of a function $\mathcal{G}_{\xi,\delta y}$
of an enlarged set of overlap parameters, which depends
on the so called Shannon transform~\cite{Tulino_RandomMatrix} of $\rho_{\xi\xi^{T}}\left(\lambda\right)$,
a quantity that measures the capacity of linear
vector channels. The resulting self-consistent equations,
which describe the statistical properties of the synaptic
weights $w_{i}$, are expressed in terms of the Stieltjes transform of
$\rho_{\xi\xi^{T}}\left(\lambda\right)$, an important tool in random matrix theory~\cite{Tao_topics}.

\begin{figure}
\centering
\includegraphics[scale=1]{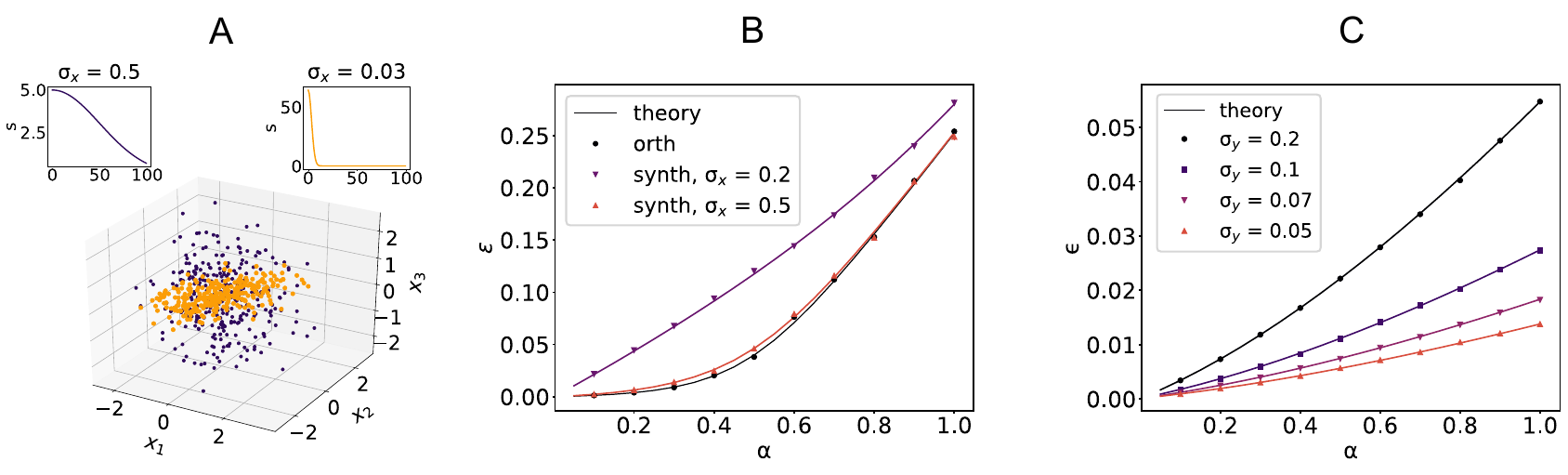}
\caption{{\bf Sample-based PCA and learning performance.}
A: First three components of inputs $\boldsymbol{\xi}_\mu$
with Gaussian singular value spectrum $s$ for two different values of $\sigma_x$
(color coded top panels). Parameters: $N=100$, $M=300$.
B: Average error $\epsilon$ for three different
singular value spectra of the input sample covariance matrix: orthogonal
model and Gaussian model with increasing $\sigma_x$ (see main text for
definition of $\sigma_x$). Outputs are i.i.d Gaussian. Parameters: $N=1000$,
$f_{E}=0.8$, $\gamma=0.1$, $\bar{x}_{E}=\bar{x}_{I}=\bar{y}=\sigma_{y}=1$, $\sigma_{I}=2\sigma_{E}=0.6$.
B: Average error $\epsilon$ for
input with orthogonal-type covariance and output $y$ with rbf-type
covariance with decreasing $\sigma_{y}$ (see main text for the definition
of $\sigma_{y}$). All remaining parameters as in A. Full lines show
analytical results. Points are averages across $50$ samples.}
\label{fig:fig_leg}
\end{figure}

We show the validity of the mean-field approach by employing two different
data models for the input signals. In the first example, valid for $\alpha\leq1$,
all the $M$ vectors $\boldsymbol{\xi}_{\mu}$ are
orthogonal to each other. This yields an eigenvalue distribution of the simple
form $\rho\left(\lambda\right)=\alpha\delta\left(\lambda-1\right)+\left(1-\alpha\right)\delta\left(\lambda\right)$,
for which the function $\mathcal{G}_{\xi,\delta y}$ can be computed explicitly \cite{Shinzato_Revisited}.
Additionally, we use a synthetic model where we explicitly
set the singular value spectrum of $\xi$ to be $s\left(\alpha\right)=\chi e^{-\frac{\alpha^{2}}{2\sigma_x^{2}}}$,
with $\chi$ a normalization factor ensuring matrix $\xi$ has unit variance.
The shape of the singular value spectrum $s$ controls the spread
of the data points $\boldsymbol{\xi}_\mu$ in the $N$-dimensional
input space, as shown in Figure \ref{fig:fig_leg}A.
As shown in Figure \ref{fig:fig_leg}B for i.i.d Gaussian output,
learning degrades as $\sigma_x$ decreases, since
inputs tend to be confined to a lower dimensional subspace 
rather than being equally distributed along input dimensions.
For $N$ large enough (in practice, for $N\gtrsim500$), the statistics
of single cases is well captured by the equations for the average case (self-averaging effect).
To get a mean-field description for a single case,
where a given input matrix $X$ is used, we further assume we have access to the
linear expansion $c_{\mu}$ of the output $y$ in the set $\lbrace \boldsymbol{v}_\mu \rbrace$ of the columns
of the $V$ matrix, namely $\boldsymbol{y}=\bar{y}+\sigma_{y}V\boldsymbol{c}$. The calculation
can be carried out in a similar way and yields, for the average regression loss, the following result:
\begin{equation}
\left<E\right>=\frac{\alpha}{2}\sigma_{y}^{2}\tilde{\Lambda}_{w}\left\langle\frac{\lambda^{y}}{\lambda^x+\tilde{\Lambda}_{w}}\right\rangle_{\lambda^x,\lambda^y}
\label{eq:energy_sample_x_y}
\end{equation}
The average in Eq~(\ref{eq:energy_sample_x_y}) is computed over the eigenvalues $\lambda^x$
of the sample covariance matrix, which correspond to the PCA variances, and $\lambda^y_\mu = c_\mu^2$ (Methods).
The quantity $\tilde{\Lambda}_w$ can be computed from a set of self-consistent equations
that link the order parameter $\Delta \tilde{q}_w$ and the first two moments of the synaptic distribution.
To better understand the role of the parameter $\tilde{\Lambda}_{w}$, it is instructive to compare
Eq~(\ref{eq:energy_sample_x_y}) with the corresponding result for unconstrained weights,
which can be derived from the pseudo-inverse solution
(Methods), $w^{*}=\left(\xi\xi^{T}+\gamma\right)^{-1}\xi y$.
The average loss is:
\begin{equation}
\left<E\right>=\frac{\alpha}{2}\sigma_{y}^{2}\gamma\left\langle \frac{\lambda^{y}}{\lambda^{x}+\gamma}\right\rangle _{\lambda^{x},\lambda^{y}}
\label{eq:energy_sample_x_y_unconstrained}
\end{equation}
Comparing Eq~(\ref{eq:energy_sample_x_y}) and Eq~(\ref{eq:energy_sample_x_y_unconstrained}),
we find that $\tilde{\Lambda}_{w}$ acts as an implicit regularization
in the sign-constrained case.  In Fig~\ref{fig:fig_leg}C, we show results
when the dimensionality of the output $y$ along the (temporal) components
of the input is modulated by taking $c\left(\alpha\right)=e^{-\frac{\alpha^{2}}{2\sigma_{y}^{2}}}$.
The perceptron performance improves as the output signals spreads
out across multiple components $\boldsymbol{v}_\mu$.
The case of i.i.d. output is recovered by taking $c_{\mu}=1$.

\section*{Discussion}
In this work, I investigated the properties of optimal solutions
of a linear perceptron with sign-constrained synapses and correlated input/output
signals, thus providing a general mean-field theory for constrained
regression in the presence of correlations. I treated both cases
where ensemble covariances are known and where
the sample covariance is given for a typical case. The latter approach,
built on a rotationally invariant assumption, allowed to link
the regression performance to the input and output statistical properties
expressed by principal component analysis.

I provided the general expression of the weight distribution for regularized regression and found that
half of the weights are set to zero, irrespectively of the fraction
of excitatory weights, provided the bias is optimized. The shape of
the synaptic distribution has been previously described in the binary
perceptron with independent input at critical capacity,
as well as in the theory of compressed sensing~\cite{Ganguli_compressed}.
I elucidated the role of the optimal bias current and its relation to
the optimal capacity and the scaling of the solution weights.
This analysis also shed light on the structural properties
of synaptic matrices which emerge when target-based methods
are used for building biologically plausible functional models of rate and spiking networks.

The theory presented in this work is relevant in the effort of
establishing quantitative comparisons between the synaptic profile of
neural circuits involved in temporal processing of dynamic signals,
such as the cerebellum~\cite{Marr_cerebellar,Wolpert_cerebellum,Herzfeld_purkinje},
and normative theories that take into account
the temporal and geometrical complexity of computational tasks.
On the other hand, the construction of progressively more biologically
plausible models of neural circuits calls for normative theories of
learning in heterogeneous networks, which can be coupled to dynamic
mean-field analysis of E-I separated circuits~\cite{KadmonSompolinsky,HarishHanselAsynchronous,mastrogiuseppeeinetworks}.

The importance of a theory of constrained regression with realistic
input/output statistics goes beyond the realm of neuroscience. Non-negativity
is commonly required to provide interpretable
results in a wide variety of inference and learning problems. Off-line
and on-line least-square estimation methods \cite{Chen_nonnegative,Nascimento_rls}
are also of great practical importance in adaptive control applications,
where constraints on the parameter range are usually imposed by 
physical plausibility.

In this work, I assumed statistical independence between inputs and
outputs. For the sake of biological plausibility, it would be interesting
to consider more general input-output correlations for regression and
binary discrimination tasks. The classical model for such
correlations is provided by the so-called teacher-student (TS) approach~\cite{Engel_statistical},
where the output $y$ is generated by a deterministic parameter-dependent
transformation of the input $x$, with a structure similar to the trained neural architecture. The problem of input/output correlations is deeply related to the issue of optimal random nonlinear expansion
both in statistical learning theory \cite{mei_generalization,Gerace_generalisation} and
theoretical neuroscience \cite{Babadi_sparseness,LitwinKumar_optimal}, with a history
dating back to the Marr-Albus theory of pattern separation in cerebellum~\cite{Cayco_patternseparation}. In a recent work, \cite{goldt_modelling} introduced
a promising generalization of TS, in which labels are generated via
a low-dimensional latent representation, and it was shown that this model
captures the training dynamics in deep networks with real world datasets.

A general analysis that fully takes into account spatio-temporal
correlations in network models could shed light on the emergence of
specific network motifs during training. In networks with non-linear
dynamics, the mathematical treatment quickly gets challenging
even for simple learning rules. In recent years, interesting work has
been done to clarify the relation between learning and network motifs,
using a variety of mean-field approaches. Examples are the study of
associative learning in spin models \cite{Brunel_cortical} and the
analysis of motif dynamics for simple learning rules in spiking networks \cite{Ocker_motifs}.
Incorporating both the temporal aspects of learning
and neural cross-correlations in E-I separated models with
realistic input/output structure is an interesting topic for future
work.

\section*{Methods}

\subsection*{Replica formalism: ensemble covariance matrix}
Using the Replica formalism~\cite{Mezard_SpinGlassBeyond}, the free energy density is written as:
\begin{equation}
-\beta f=\frac{1}{N}\lim_{n\to0}\frac{\partial}{\partial n}\log\left\langle Z^{n}\right\rangle_{x,y}
\label{eq:app_free_energy}
\end{equation}
The function $Z^{n}$ can be computed by considering a finite number
$n$ of replicas of the vector $\boldsymbol{w}$ and subsequently
taking a continuation $n \in \mathbb{R}$. In the large $N$ limit, $f$ can
be written as the sum of two contributions $-\beta f=\mathcal{G_{S}}+\alpha\mathcal{G}_{E}$,
respectively called \emph{entropic} and \emph{energetic} part, which
depend on a small set of \emph{order parameters}, to be determined
by solving the saddle point equations arising from the expression
$\left\langle Z^{n}\right\rangle _{x,y}= e^{-\beta Nf}$ in the limit $n\to 0$. In the
following, we will usually drop the subscript in the average $\left\langle \cdot\right\rangle _{x,y}$.
To simplify the formulas, we introduce the $\mathcal{O}\left(1\right)$
weights $J_{i}=\sigma_{i}\sqrt{N}w_{i}$. In terms of these rescaled
variables, the loss function in Eq~(\ref{eq:loss_function}) takes the
form:
\begin{equation}
E\left(\boldsymbol{w};\gamma,\xi,y\right)=\frac{1}{2}\sum_{\mu=1}^{M}\left(\sum_{i=1}^{N}\frac{J_{i}}{\sqrt{N}}\xi_{i\mu}+\frac{1}{\sqrt{N}}\sum_{i=1}^{N}\frac{\bar{x}_{i}}{\sigma_{i}}J_{i}+I\sqrt{N}-y_{\mu}\right)^{2}+\frac{\gamma}{2}\sum_{i=1}^N\frac{J_{i}^{2}}{\sigma_{i}^{2}}\label{eq:app_loss_function}
\end{equation}
by virtue of $x_{i\mu}=\bar{x}_{i}+\sigma_{i}\xi_{i\mu}$.
We proceed by inserting the definitions $M^{a}=\frac{1}{\sqrt{N}}\sum_{i=1}^{N}\frac{\bar{x}_{i}}{\sigma_{i}}J_{i}+I\sqrt{N}$
and $\Delta_{\mu a}=\sum_{i=1}^N\xi_{i\mu}\frac{J_{ia}}{\sqrt{N}}-\sigma_{y}\delta y_{\mu}$
with the aid of appropriate $\delta$ functions. Assuming \emph{balance},
namely $M^{a}=\bar{y}$, the averaged replicated partition function
$\left\langle Z^{n}\right\rangle $ is:
\begin{align}
\left\langle Z^{n}\right\rangle  & =\int\prod_{a}d\mu\left(\boldsymbol{w}_{a}\right)\int\prod_{\mu a}\frac{d\Delta_{\mu a}du_{\mu a}}{2\pi}Z_{\xi,\delta y}\nonumber \\
 & e^{\sum_{a}\hat{M}^{a}\left(\sqrt{N}M^{a}-\sum_{i}\frac{\bar{x}_{i}}{\sigma_{i}}J_{i}-NI\right)-i\sum_{\mu a}u_{\mu a}\Delta_{\mu a}-\frac{\beta}{2}\sum_{\mu a}\Delta_{\mu a}^{2}-\frac{\beta\gamma}{2}\sum_{ia}\frac{J_{ia}^{2}}{\sigma_{i}^{2}}}\label{eq:app_replicated_Z}
\end{align}
where:
\begin{equation}
Z_{\xi,\delta y}=\left\langle e^{i\sum_{\mu a}u_{\mu a}\left(\sum_{i}\xi_{i\mu}\frac{J_{ia}}{\sqrt{N}}-\sigma_{y}\delta y_{\mu}\right)}\right\rangle _{\xi,\delta y}\label{eq:app_cumulant_generating_covariance}
\end{equation}
The calculation can be carried out by introducing overlap order parameters
$Nq_{w}^{ab}=\sum_{i=1}^{N}J_{ia}J_{ib}$ with the use of $n\left(n-1\right)/2$
additional $\delta$ functions, together with their conjugate variables
$\hat{q}^{ab}_w$. Owing to the convexity of the regression problem,
we use a Replica Symmetry (RS) \cite{Mezard_SpinGlassBeyond} ansatz
$q_{w}^{ab}=q_{w}+\delta_{ab}\Delta q_{w}$, and correspondingly for
the conjugate parameters. Additionally, we will take $M^{a}=M$ and $\hat{M}^{a}=\hat{M}$.

\subsubsection*{Entropic part}
The total volume of configurations $\boldsymbol{w}_a$ for fixed values of the overlap
parameters is given by the \emph{entropic part}, which can be computed
at RS level by standard methods, yielding:
\begin{align}
	\mathcal{G}_{S} & =\sum_{c\in\left\{ E,I\right\} }f_{c}\int Dz\log\int_{0}^{\infty}dJe^{-\frac{J^{2}}{2}\left(\Delta\hat{q}_{w}+\frac{\beta\gamma}{\sigma_{c}^2}\right)+s_{c}\left(z\sqrt{\hat{q}_{w}}-\eta_{c}\hat{M}\right)J}\nonumber\\
	 & -\hat{M}I+\frac{\Delta\hat{q}_{w}}{2}\left(\Delta q_{w}+q_{w}\right)-\frac{\hat{q}_{w}\Delta q_{w}}{2}
\label{eq:app_entropic_part}
\end{align}
where $Dz=\frac{e^{-\frac{z^{2}}{2}}}{\sqrt{2\pi}}$. In Eq~(\ref{eq:app_entropic_part}),
we introduced the notations $f_I=1-f_E$ and $s_{E}=-s_{I}=1$.

\subsubsection*{Energetic part}
In order to compute the\emph{ energetic }part, we first need to evaluate
the average with respect to $\xi$ and $\delta y$ in Eq~(\ref{eq:app_cumulant_generating_covariance}).
Performing the two Gaussian integrals we get:
\begin{equation}
Z_{\xi,\delta y}=e^{-\frac{1}{2}\sum_{\mu\nu}\sum_{ab}q_{w}^{ab}u_{\mu a}u_{\nu b}C_{\mu\nu}^{x}-\frac{\sigma_{y}^{2}}{2}\sum_{\mu\nu}\sum_{ab}u_{\mu a}u_{\nu b}C_{\mu\nu}^{y}}\label{eq:app_generating_C}
\end{equation}
from which: 
\begin{align}
e^{\alpha N\mathcal{G}_{E}} & \propto\int\prod_{\mu a}\frac{d\Delta_{\mu a}du_{\mu a}}{2\pi}e{}^{-\frac{\beta}{2}\sum_{\mu a}\Delta_{\mu a}^{2}-\frac{1}{2}\sum_{\mu\nu}\sum_{ab}q_{w}^{ab}u_{\mu a}u_{\nu b}C_{\mu\nu}^{x}}\nonumber\\
 & e^{-\frac{\sigma_{y}^{2}}{2}\sum_{\mu\nu}\sum_{ab}u_{\mu a}u_{\nu b}C_{\mu\nu}^{y}-i\sum_{\mu a}u_{\mu a}\Delta_{\mu a}}
\end{align}
In the special case $C^{x}=C^{y}\equiv C$, we can jointly rotate
the $u$'s and $\Delta$'s variables, using $C=V\Lambda V^{T}$, to
obtain:
\begin{align}
e^{\alpha N\mathcal{G}_{E}} & \propto\int\prod_{\mu a}\frac{d\Delta_{\mu a}du_{\mu a}}{2\pi}e^{-\frac{\beta}{2}\sum_{\mu a}\Delta_{\mu a}^{2}-\frac{1}{2}\sum_{\mu}\sum_{ab}q_{w}^{ab}u_{\mu a}u_{\mu b}\lambda_{\mu}}\nonumber\\
 & e^{-\frac{\sigma_{y}^{2}}{2}\sum_{\mu}\sum_{ab}u_{\mu a}u_{\mu b}\lambda_{\mu}-i\sum_{\mu a}u_{\mu a}\Delta_{\mu a}}
\end{align}
Within the RS ansatz, we get:
\begin{equation}
\mathcal{G}_{E}=-\frac{1}{2}\left\langle \log\left(1+\beta\Delta q_{w}\lambda\right)\right\rangle _{\lambda}-\frac{\beta}{2}\left(q_{w}+\sigma_{y}^{2}\right)\left\langle \frac{\lambda}{1+\beta\Delta q_{w}\lambda}\right\rangle _{\lambda}\label{eq:app_energetic_part_cov_only_x}
\end{equation}
The brakets $\left\langle \right\rangle _{\lambda}$ in Eq~(\ref{eq:app_energetic_part_cov_only_x})
stand for an average over the eigenvalue distribution $\rho\left(\lambda\right)$
of $C$ in the $N\to\infty$ limit, assuming self-averaging \cite{Monasson_properties,Tarkowski_optimal}.
When $C^{x}\neq C^{y}$, we can derive a similar expression under
the assumption of a ring topology in pattern space (corresponding to period boundary
conditions in the index $\mu$). In the main text, we show that the expression
\begin{equation}
\alpha \mathcal{G}_{E}=-\frac{1}{2N}\sum_{\mu}\log\left(1+\beta\Delta q_w\lambda_{\mu}^{x}\right)-\frac{\beta}{2N}\sum_{\mu}\frac{q_w\lambda_{\mu}^{x}+\sigma^2_y\lambda_{\mu}^{y}}{1+\beta\Delta q_w\lambda_{\mu}^{x}}\label{eq:app_energetic_part_cov_xy}
\end{equation}
yields good results also when $C^{x}$ and $C^{y}$ are covariance matrices
of stationary discrete-time processes.

\subsubsection*{Saddle point equations}
All in all, the free-energy is:
\begin{align}
	-\beta f & =-\hat{M}I+\frac{\Delta\hat{q}_{w}}{2}\left(\Delta q_{w}+q_{w}\right)-\frac{\hat{q}_{w}\Delta q_{w}}{2}\nonumber \\
	 & -\frac{1}{2N}\sum_{\mu}\log\left(1+\beta\Delta q_{w}\lambda_{\mu}^{x}\right)-\frac{\beta}{2N}\sum_{\mu}\frac{q_{w}\lambda_{\mu}^{x}+\sigma_{y}^{2}\lambda_{\mu}^{y}}{1+\beta\Delta q_{w}\lambda_{\mu}^{x}}+\nonumber \\
	  & \sum_{c\in\left\{ E,I\right\} }f_{c}\int Dz\log\int_{0}^{\infty}dJe^{-\frac{J^{2}}{2}\left(\Delta\hat{q}_{w}+\frac{\beta\gamma}{\sigma_{c}^2}\right)+s_{c}\left(z\sqrt{\hat{q}_{w}}-\eta_{c}\hat{M}\right)J}
\end{align}
The equations stemming from the entropic part can be written as:
\begin{align}
 & q_{w}=\left\langle \left\langle J\right\rangle _{J}^{2}\right\rangle _{z}\label{eq:app_saddle_q_w}\\
 & \Delta q_{w}=\left\langle \left\langle J^{2}\right\rangle _{J}\right\rangle _{z}-\left\langle \left\langle J\right\rangle _{J}^{2}\right\rangle _{z}\label{eq:app_saddle_delta_q_w}\\
& I+\sum_{c\in\left\{ E,I\right\} }\eta_{c}\left\langle \left\langle J\right\rangle _{J}\right\rangle _{z}=0
\label{eq:app_saddle_b}
\end{align}
where the averages $\left\langle \,\right\rangle _{J}$ and
$\left\langle \,\right\rangle _{z}$ in Eq~(\ref{eq:app_saddle_q_w}),
(\ref{eq:app_saddle_delta_q_w}), (\ref{eq:app_saddle_b}) are taken with
respect to the mean-field distribution of the $J$ weights:
\begin{align}
p\left(J;z\right)&	\propto\sum_{c\in\left\{ E,I\right\} }f_{c}p_c\left(J;z\right)\\
p_c\left(J;z\right)&	\propto\theta\left(s_{c}J\right)e^{-\frac{J^{2}}{2}\left(\Delta\hat{q}_{w}+\frac{\beta\gamma}{\sigma_{c}}\right)+J\left(z\sqrt{\hat{q}_{w}}-\eta_{c}\hat{M}\right)}
\end{align}
where $z$ is a standard normal variable and $\theta$ is the Heaviside function: $\theta\left(x\right)=1$ when $x>0$ and $0$ otherwise.
In the $\beta\to\infty$ limit, the unicity of solution for $\gamma>0$
implies that $\Delta q_{w}\to0$. We therefore use the following scalings
for the order parameters:
\begin{align}
 & \beta\Delta q_{w}=\Delta\tilde{q}_{w}\label{eq:app_scaling_delta_q_w}\\
 & \hat{q}_{w}=\beta^{2}C\label{eq:app_scaling_q_hat_w}\\
 & \Delta\hat{q}_{w}=\beta A\label{eq:app_scaling_delta_q_hat_w}\\
 & \hat{M}=\beta B\sqrt{C}\label{eq:app_scaling_m_hat}
\end{align}
while $q_{w}=\mathcal{O}\left(1\right)$. In this scaling, Eq~(\ref{eq:app_saddle_q_w}),
(\ref{eq:app_saddle_delta_q_w}), (\ref{eq:app_saddle_b}) take the form:
\begin{align}
 & \frac{q_{w}}{C}=\sum_{c\in \left\{E,I\right\}}\frac{f_{c}}{\left(A+\frac{\gamma}{\sigma_{c}^{2}}\right)^{2}}\left(\left(1+\eta_{c}^{2}B^{2}\right)H\left(s_c\eta_{c}B\right)-s_c\eta_{c}BG\left(\eta_{c}B\right)\right)\label{eq:app_saddle_entropic_q_w}\\
 & \Delta\tilde{q}_{w}=\sum_{c\in \left\{E,I\right\}}\frac{f_{c}}{A+\frac{\gamma}{\sigma_{c}^{2}}}H\left(s_c\eta_{c}B\right)\label{eq:app_saddle_entropic_delta_qt_w}\\
& \frac{I}{\sqrt{C}}=\sum_{c\in \left\{E,I\right\}}\frac{f_{c}}{A+\frac{\gamma}{\sigma_{c}^{2}}}\left(\eta_{c}^{2}BH\left(s_{c}\eta_{c}B\right)-s_{c}\eta_{c}G\left(\eta_{c}B\right)\right)\label{eq:app_saddle_entropic_I}
\end{align}
where $G\left(z\right)=\frac{e^{-\frac{z^{2}}{2}}}{\sqrt{2\pi}}$.
The squared norm of the weights $v=\sum_{i=1}^N w_{i}^{2}$ is given by $v=2\partial_{\gamma}f$:
\begin{equation}
v=C\sum_{c\in \left\{E,I\right\}}\frac{f_{c}}{\sigma_{c}^{2}\left(A+\frac{\gamma}{\sigma_{c}^{2}}\right)^{2}}\left(\left(1+\eta_{c}^{2}B^{2}\right)H\left(s_c\eta_{c}B\right)-s_c\eta_{c}BG\left(\eta_{c}B\right)\right)\label{eq:app_saddle_entropic_norm}
\end{equation}
In the $\beta\to\infty$, it can be easily shown that the mean-field weight probability density of the rescaled
weights $\sqrt{N}w_{i}$ is a superposition of a $\delta$ function in zero
and two truncated Gaussian densitites: 
\begin{equation}
p\left(\sqrt{N}w\right)=p_{0}\left(B\right)\delta\left(w\right)+\sum_{c}f_{c}G\left(\sqrt{N}w;M_{c},\Sigma_{c}\right)\theta\left(s_{c}J\right)\label{eq:app_distribution}
\end{equation}
where the mean and standard deviation of the Gaussians $G\left(\cdot;M,\Sigma\right)$
are:
\begin{align}
 & M_{c}=-\frac{\eta_{c}B\sqrt{C}}{\sigma_c A + \frac{\gamma}{\sigma_c}},\\
 & \Sigma_{c}=\frac{\sqrt{C}}{\sigma_c A + \frac{\gamma}{\sigma_c}}
\end{align}
The fraction of zero weights is given by:
\[
p_{0}\left(B\right)=f_{E}H\left(-\eta_{E}B\right)+\left(1-f_{E}\right)H\left(\eta_{I}B\right)
\]
where $H\left(x\right)=\int_{x}^{\infty}Dz$.
The two remaining saddle point equations are:
\begin{align}
 & C=\frac{1}{N}\sum_{\mu}\lambda_{\mu}^{x}\frac{q_w\lambda_{\mu}^{x}+\sigma_{y}^{2}\lambda_{\mu}^{y}}{\left(1+\Delta\tilde{q}_w\lambda_{\mu}^{x}\right)^{2}}\label{eq:app_saddle_energetic_C}\\
 & A=\frac{1}{N}\sum_{\mu}\frac{\lambda_{\mu}^{x}}{1+\Delta\tilde{q}_w\lambda_{\mu}^{x}}\label{eq:app_saddle_energetic_A}
\end{align}
Optimizing $f$ with respect to the bias $b=I\sqrt{N}$ immediately
implies $B=0$, by virtue of Eq~(\ref{eq:app_scaling_m_hat}). Using
the scaling assumptions Eq~(\ref{eq:app_scaling_delta_q_w})-(\ref{eq:app_scaling_m_hat})
together with the saddle point Eq~(\ref{eq:app_saddle_entropic_delta_qt_w})-(\ref{eq:app_saddle_energetic_A}),
we get Eq~(\ref{eq:delta_q_tilde}) in the main text, that is valid for any $\alpha$ for $\gamma>0$. In the
unregularized case ($\gamma=0$), it describes solutions in the error regime $\alpha>\alpha_c$.
The optimal bias $b$ can be computed by $I\sqrt{N}$ using Eq~(\ref{eq:app_saddle_entropic_I}),
that is valid up to the an $\mathcal{O}\left(1\right)$ term equal to $\bar{y}$ (Fig~\ref{fig:fig3}B).
The expression for the average minimal energy is:
\begin{equation}
\left<E\right>=\frac{\sigma_{y}^{2}}{2N}\sum_{\mu}\frac{\lambda_{\mu}^{y}}{1+\Delta\tilde{q}_w\lambda_{\mu}^{x}}
\label{eq:app_energy_cov}
\end{equation}

\subsubsection*{Spectrum of exponential and rbf covariance}
For the exponential covariance $C_{\mu\nu}=e^{-\frac{\left|\mu-\nu\right|}{\tau}}$
one has:

\[
\lambda\left(\phi\right)=\frac{1-x^{2}}{1-2x\cos\phi+x^{2}}
\]
with $x=e^{-\frac{1}{\tau}}$. In
the rbf case $C_{\mu\nu}=e^{-\frac{\left|\mu-\nu\right|^{2}}{2\tau^{2}}}$,
the spectrum is:
\[
\lambda\left(\phi\right)=\vartheta_{3}\left(\frac{\phi}{2},e^{-\frac{1}{2\tau^{2}}}\right)
\]
with $\vartheta_{3}\left(z,q\right)=1+2\sum_{n=1}^{\infty}q^{n^{2}}\cos\left(2nz\right)$
the Jacobi theta function of 3rd type.

\subsection*{Replica formalism: sample covariance matrix}
In the case of a sample covariance matrix, the free-energy is a sum of three contributions $-\beta f=\mathcal{G_{S}}+\mathcal{G}_{\xi,\delta y}+\alpha\mathcal{G}_{E}$.
The entropic part is unchanged. As explained in the main text, we
extend the calculations in \cite{Shinzato_learning,Shinzato_Revisited}
to the case where the linear expansion of $y_\mu$ on the right singular
vectors $V_{\cdot\mu}$ is known, by taking $\delta y_{\mu}=\sum_{\nu}V_{\mu\nu}c_{\nu}$.
Using again the expressions $\left(X\right)_{i\mu}=\bar{x}_{i}+\sigma_{i}\xi_{i\mu}$
and $\xi=USV^{T}$, the replicated cumulant generating function for
the joint (mean-removed) input and output is:
\begin{align}
Z_{\xi,\delta y} & =\left\langle \exp\left(i\sum_{a}\tilde{\boldsymbol{J}}_{a}^{T}S\tilde{\boldsymbol{u}}_{a}-i\sigma_{y}\boldsymbol{c}^{T}\sum_{a}\tilde{\boldsymbol{u}}_{a}\right)\right\rangle _{p\left(\tilde{\boldsymbol{J}}_{a},\tilde{\boldsymbol{u}}_{a}\right)}\label{eq:app_cumulant_generating_sample}
\end{align}
where we used the change of variables $\tilde{J}_{ia}=\sum_{k}U_{ki}J_{ka}$
and $\tilde{u}_{\mu a}=\sum_{k}V_{k\mu}u_{k a}$. The average in
Eq~(\ref{eq:app_cumulant_generating_sample}) is taken over the joint
distribution $p\left(\tilde{\boldsymbol{J}}_{a},\tilde{\boldsymbol{u}}_{a}\right)$
resulting from averaging over the Haar measure on the orthogonal matrices $U$ and
$V$. For a single replica, $Z_{\xi,\delta y}$ will
only depend on the squared norms $Q_{w}=\sum_{i}\frac{\tilde{J}_{i}^{2}}{N}$
and $Q_{u}=\sum_{\mu}\frac{\tilde{u}_{\mu}^{2}}{M}$ of the two vectors
$\tilde{\boldsymbol{J}}$ and $\tilde{\boldsymbol{u}}$. We can therefore
write the average in the following way:
\begin{equation}
\left\langle \exp\left(i\tilde{\boldsymbol{J}}^{T}S\tilde{\boldsymbol{u}}-i\sigma_{y}\boldsymbol{c}^{T}\tilde{\boldsymbol{u}}\right)\right\rangle _{p\left(\tilde{\boldsymbol{J}},\tilde{\boldsymbol{u}}\right)}\propto\int\delta\left(\left|\tilde{\boldsymbol{J}}\right|^{2}-NQ_{w}\right)\delta\left(\left|\tilde{\boldsymbol{u}}\right|^{2}-MQ_{u}\right)e^{i\tilde{\boldsymbol{J}}^{T}S\tilde{\boldsymbol{u}}-i\sigma_{y}\boldsymbol{c}^{T}\tilde{\boldsymbol{u}}}\label{eq:app_generating_svd}
\end{equation}
Introducing Fourier representation for the $\delta$ functions, we are
left with an expression involving an $N+M$ dimensional Gaussian integral:
\begin{align}
 & \int\frac{d\Lambda_{w}}{4\pi i}\frac{d\Lambda_{u}}{4\pi i}e^{\frac{N\Lambda_{w}Q_{w}}{2}+\frac{M\Lambda_{u}Q_{u}}{2}}\int d\tilde{\boldsymbol{J}}d\tilde{\boldsymbol{u}}e^{-\frac{\Lambda_{w}}{2}\left|\tilde{\boldsymbol{J}}\right|^{2}-\frac{\Lambda_{u}}{2}\left|\tilde{\boldsymbol{u}}\right|^{2}+i\tilde{\boldsymbol{J}}^T S \tilde{\boldsymbol{u}}-i\sigma_{y}\boldsymbol{c}^{T}\tilde{\boldsymbol{u}}}\nonumber \\
 & =\frac{\left(2\pi\right)^{\frac{N+M}{2}}}{\left(4\pi i\right)^{2}}\int d\Lambda_{w}d\Lambda_{u}e^{\frac{N\Lambda_{w}Q_{w}}{2}+\frac{M\Lambda_{u}Q_{u}}{2}}\det\mathcal{M}{}^{-\frac{1}{2}}\exp\left(-\frac{\sigma_{y}^{2}}{2}\left(\begin{array}{cc}
\boldsymbol{0} & \boldsymbol{c}\end{array}\right)\mathcal{M}^{-1}\left(\begin{array}{c}
\boldsymbol{0}\\
\boldsymbol{c}
\end{array}\right)\right)\label{eq:app_F_integral}
\end{align}
where 
\[
\mathcal{M}=\left(\begin{array}{cc}
\Lambda_{w}{\1}_{N} & -iS\\
-iS^{T} & \Lambda_{u}{\1}_{M}
\end{array}\right)
\]
and ${\1}_{K}$ is the identity matrix of dimension $K$. Following \cite{Shinzato_Revisited}, the determinant
can be easily calculated:
\begin{align}
\frac{1}{N}\log\det\mathcal{M} & =\frac{1}{N}\sum_{k=1}^{\min\left(N,M\right)}\log\left(\lambda_{k}^{x}+\Lambda_{w}\Lambda_{u}\right)+\frac{\left(N-\min\left(N,M\right)\right)}{N}\log\Lambda_{u}\to\\
 & \to \left\langle \log\left(\lambda_{k}^{x}+\Lambda_{w}\Lambda_{u}\right)\right\rangle _{\lambda^{x}}+\left(\alpha-1\right)\log\Lambda_{u}
\end{align}
where the limit is taken for $N\to\infty$ and the average is with respect to the eigenvalue distribution $\rho\left(\lambda^{x}\right)$.
As for the quadratic portion of the Gaussian integral, calling $\lambda_{k}^{y}=c_{k}^{2}$,
we will use the shorthand 
\[
\left\langle \frac{\lambda^{y}}{\lambda^{x}+\Lambda_{w}\Lambda_{u}}\right\rangle _{\lambda^{x},\lambda^{y}}\equiv\frac{\Lambda_{w}}{M}\sum_{k=1}^{\Omega}\frac{\lambda_{k}^{y}}{\lambda^{x}_k+\Lambda_{w}\Lambda_{u}}+\frac{\left(M-\Omega\right)}{M}\sum_{k=M+1}^{\Omega}\frac{\lambda_{k}^{y}}{\Lambda_{w}}
\]
where $\Omega=\max\left(N,M\right)$.
Considering now the replicated generating function, all the $n\left(2n-1\right)$
cross-product $\boldsymbol{J}_{a}\cdot\boldsymbol{J}_{b}=\tilde{\boldsymbol{J}}_{a}\cdot\tilde{\boldsymbol{J}}_{b}$
and $\boldsymbol{u}_{a}\cdot\boldsymbol{u}_{b}=\tilde{\boldsymbol{u}}_{a}\cdot\tilde{\boldsymbol{u}}_{b}$
must be conserved via the multiplication of $U$ and $V$. Together with the overlap
parameters $Nq^{ab}_w=\sum_{i}J_{ia}J_{ib}$, we additionally introduce
the quantities $Mq_{u}^{ab}=\sum_{\mu}u_{\mu a}u_{\mu b}$. In the RS
case, we again take: $q_{w}^{ab}=q_{w}+\delta_{ab}\Delta q_{w}$ and,
similarly for the $u$'s, $q_{u}^{ab}=-q_{u}+\delta_{ab}\Delta q_{u}$.
In the basis where both $q_{w}^{ab}$ and $q^{ab}_u$ are diagonal, the expression
becomes
\begin{equation}
Z_{\xi,\delta y}=\left<e^{i\tilde{\boldsymbol{J}}_{1}^{T}S\tilde{\boldsymbol{u}}_{1}-i\sigma_{y}\boldsymbol{c}^{T}\sqrt{n}\tilde{\boldsymbol{u}}_{1}}\prod_{b=2}^{n}e^{i\tilde{\boldsymbol{J}}_{b}^{T}S\tilde{\boldsymbol{u}}_{b}}\right>
\end{equation}
so in the limit $n\to0$ we have:
\begin{equation}
2\mathcal{G}_{\xi,\delta y}=F\left(\Delta q_{w},\Delta q_{u}\right)+q_{w}\frac{\partial F\left(\Delta q_{w},\Delta q_{u}\right)}{\partial\Delta q_{w}}-q_{u}\frac{\partial F\left(\Delta q_{w},\Delta q_{u}\right)}{\partial\Delta q_{u}}-\alpha\sigma_{y}^{2}K\left(\Lambda_{w},\Lambda_{u}\right)\label{eq:app_F_replicated}
\end{equation}
with the function $F$ given by:
\begin{align}
F\left(x,y\right) & =\Extr_{\Lambda_{w},\Lambda_{u}}\left\{ -\left\langle \log \left(\lambda^{x}+\Lambda_{w}\Lambda_{u}\right)\right\rangle _{\lambda^{x}}-\left(\alpha-1\right)\log\Lambda_{u}+\Lambda_{w}x+\alpha\Lambda_{u}y\right\}\nonumber\\
 & -\log x-\alpha\log y-\left(1+\alpha\right)
 \label{eq:app_F}
\end{align}
and $K\left(\Lambda_{w},\Lambda_{u}\right)=\Lambda_{w}\left\langle \frac{\lambda^{y}}{\lambda^x+\Lambda_{w}\Lambda_{u}}\right\rangle _{\lambda^{x},\lambda^{y}}$.
In Eq~(\ref{eq:app_F_replicated}), it is intended that $\Lambda_{w}$
and $\Lambda_{w}$ are implied by the Legendre Transform conditions:
\begin{align}
 & \Delta q_{w}=\Lambda_{u}\left\langle \frac{1}{\lambda^{x}+\Lambda_{w}\Lambda_{u}}\right\rangle _{\lambda^{x}}\label{eq:app_legendre_lambda_w}\\
 & \alpha\Delta q_{u}=\frac{\alpha-1}{\Lambda_{u}}+\Lambda_{w}\left\langle \frac{1}{\lambda^{x}+\Lambda_{w}\Lambda_{u}}\right\rangle _{\lambda^{x}}\label{eq:app_legendre_lambda_u}
\end{align}
The calculation of the energetic part $\mathcal{G}_{E}$ is standard
and gives:
\begin{equation}
2\mathcal{G}_{E}=\Delta\hat{q}_{u}\left(\Delta q_{u}-q_{u}\right)+\hat{q}_{u}\Delta q_{u}-\log\left(1+\beta\Delta\hat{q}_{u}\right)-\beta\frac{\hat{q}_{u}}{1+\beta\Delta\hat{q}_{u}}\label{eq:app_energetic_part_sample_cov}
\end{equation}
Eliminating $\hat{q}_{u}$ and $\Delta\hat{q}_{u}$ at the saddle
point in Eq~(\ref{eq:app_energetic_part_sample_cov}), $\mathcal{G}_{E}$ reduces to:
\begin{equation}
\mathcal{G}_{E}=\frac{q_{u}-\Delta q_{u}}{2\beta}-\frac{q_{u}}{2\Delta q_{u}}+\frac{1}{2}\log\Delta q_{u}\label{eq:app_energetic_part_sample_cov_simplified}
\end{equation}

\subsubsection*{Saddle point equations}
The entropic saddle point equations are unchanged. The final expression
$-\beta f=\mathcal{G_{S}}+\mathcal{G}_{\xi,\delta y}+\alpha\mathcal{G}_{E}$
implies the following saddle point equations:
\begin{align}
 & \Delta\hat{q}_{w}+\frac{\partial F}{\partial\Delta q_{w}}=0\label{eq:app_saddle_delta_q_hat_w}\\
 & \frac{\alpha}{\Delta q_{u}}-\frac{\alpha}{\beta}+\frac{\partial F}{\partial\Delta q_{u}}=0\label{eq:app_saddle_delta_q_u}\\
 & \hat{q}_{w}=q_{w}\frac{\partial^{2}F}{\partial\Delta q_{w}^{2}}-q_{u}\frac{\partial^{2}F}{\partial\Delta q_{w}\Delta q_{u}}-\alpha\sigma_{y}^{2}\frac{\partial K}{\partial\Delta q_{w}}\label{eq:app_saddle_q_hat_w}\\
 & \alpha\frac{q_{u}}{\Delta q_{u}^{2}}=q_{u}\frac{\partial^{2}F}{\partial\Delta q_{u}^{2}}-q_{w}\frac{\partial^{2}F}{\partial\Delta q_{w}\Delta q_{u}}+\alpha\sigma_{y}^{2}\frac{\partial K}{\partial\Delta q_{u}}\label{eq:app_saddle_q_u}
\end{align}
The saddle point values of the conjugate Legendre variables $\Lambda_w$, $\Lambda_u$
greatly simplify the expression for the first and second derivatives
of $F$. Indeed, from Eq~(\ref{eq:app_saddle_delta_q_hat_w}), (\ref{eq:app_saddle_delta_q_u}) one has:
\begin{align}
 & \Lambda_{w}=\frac{1}{\Delta q_{w}}-\Delta\hat{q}_{w}\label{eq:app_saddle_lambda_w}\\
 & \Lambda_{u}=\beta^{-1}\label{eq:app_saddle_lambda_u}
\end{align}
or, setting $\Lambda_{w}=\beta\tilde{\Lambda}_{w}$:
\begin{equation}
\tilde{\Lambda}_{w}=\frac{1}{\Delta\tilde{q}_{w}}-A\label{eq:app_saddle_delta_qt_w_lambda_w}
\end{equation}
In particular, Eq~(\ref{eq:app_legendre_lambda_w}) shows that $\Delta \tilde{q}_w$ is expressed
by a Stieltjes transform of $\rho\left(\lambda^x\right)$ and the first term in
Eq~(\ref{eq:app_F}) is its Shannon transform.
In the limit $\beta\to\infty$, using the following additional scaling
relations for the $u$ overlaps:
\begin{align}
 & q_{u}=\beta^{2}\tilde{q}_{u}\label{eq:app_scaling_q_u}\\
 & \Delta q_{u}=\beta\Delta\tilde{q}_{u}\label{eq:app_scaling_delta_q_u}
\end{align}
we get the expression for the energy:
\[
\left<E\right>=\frac{\alpha}{2}\sigma_{y}^{2}\tilde{\Lambda}_{w}\left\langle \frac{\lambda^{y}}{\lambda^{x}+\tilde{\Lambda}_{w}}\right\rangle _{\lambda^x,\lambda^y}
\]

\subsubsection*{i.i.d. and unconstrained cases}
Either setting $K=0$ of $\lambda^{y}=0$ reverts back to the i.i.d. output case. In the special case of i.i.d. inputs, the eigenvalue
distribution is Marchenko-Pastur
\begin{equation}
\rho\left(\lambda\right)=\frac{\text{\ensuremath{\sqrt{\left(\lambda-\lambda_{-}\right)\left(\lambda_{+}-\lambda\right)}}}}{2\pi\lambda}
\end{equation}
with $\lambda_{+/-}=\left(1\pm\sqrt{\alpha}\right)^{2}$, from which
$F\left(\Delta q_{w},\Delta q_{u}\right)=-\frac{\alpha}{2}\Delta q_{w}\Delta q_{u}$.
The saddle point equations are essentially the same as the ones in
the previous section with $C_{\mu\nu}^{x}=C_{\mu\nu}^{y}=\delta_{\mu\nu}$.

Let us also note that, in the simple unconstrained case, taking for
simplicity $\bar{x}_{i}=0$ and $b=0$, the entropic part can be worked
out to be, up to constant terms:
\begin{equation}
2\mathcal{G}_{S}=\log\Delta q_{w}+\frac{q_{w}}{\Delta q_{w}}-\beta\gamma\left(\Delta q_{w}+q_{w}\right)
\end{equation}
which, at the saddle point, implies $\tilde{\Lambda}_{w}=\gamma$.
The mean-field distribution $p\left(\sqrt{N}w\right)$ is a zero-mean Gaussian
with variance $v=q_{w}$. Using the properties of the Hessian of the
Legendre Transform, it is easy to show that:
\begin{align}
 & q_{w}=\alpha\frac{\partial K}{\partial\Lambda_{w}}=\alpha\left\langle \frac{\lambda^{x}\lambda^{y}}{\left(\lambda^{x}+\gamma\right)^{2}}\right\rangle _{\lambda^{x},\lambda^{y}}\\
 & \left<E\right>=\frac{\alpha}{2}\sigma_{y}^{2}\gamma\left\langle \frac{\lambda^{y}}{\lambda^{x}+\gamma}\right\rangle _{\lambda^{x},\lambda^{y}}
\end{align}
These expressions can also be derived from the pseudo-inverse solution (we take $\bar{y}=0$ for simplicity)
$w^{*}=\left(\xi\xi^{T}+\gamma\right)^{-1}\xi y$ by taking an average
across $\xi$ and $y$ in the two expressions:
\begin{align}
 & v=\left<w^{*T}w^{*}\right>=\Tr\left(\xi yy^{T}\xi^{T}\left(\xi\xi^{T}+\gamma\right)^{-2}\right)\\
 & \left<E\right>= \frac{1}{2} \left<y^{T}y\right> - \frac{1}{2} \Tr\left(\xi yy^{T}\xi^{T}\left(\xi\xi^{T}+\gamma\right)^{-1}\right)
\end{align}
The i.i.d. output case also follows by performing independent averages
over $y$ and $\xi$. 

\begin{acknowledgments}
The author would like to thank L.F. Abbott and Francesco Fumarola for constructive criticism of the manuscript.
\end{acknowledgments}

\bibliographystyle{unsrt}
\bibliography{references_arxiv}

\begin{thebibliography}{10}

\bibitem{Song_training}
H.~Francis Song, Guangyu~R. Yang, and Xiao-Jing Wang.
\newblock Training excitatory-inhibitory recurrent neural networks for
  cognitive tasks: A simple and flexible framework.
\newblock {\em PLOS Computational Biology}, 12(2):1--30, 02 2016.

\bibitem{NicolaClopathSupervised}
Wilten Nicola and Claudia Clopath.
\newblock Supervised learning in spiking neural networks with force training.
\newblock {\em Nature Communications}, 8(1):2208, 2017.

\bibitem{Ingrosso_training}
Alessandro Ingrosso and L.~F. Abbott.
\newblock Training dynamically balanced excitatory-inhibitory networks.
\newblock {\em PLOS ONE}, 14(8):1--18, 08 2019.

\bibitem{kimlearning}
Christopher~M Kim and Carson~C Chow.
\newblock Learning recurrent dynamics in spiking networks.
\newblock {\em eLife}, 7:e37124, Sep 2018.

\bibitem{Deneve_spikebyspike}
Wieland Brendel, Ralph Bourdoukan, Pietro Vertechi, Christian~K. Machens, and
  Sophie Den\`eve.
\newblock Learning to represent signals spike by spike.
\newblock {\em PLOS Computational Biology}, 16(3):1--23, 03 2020.

\bibitem{Brunel_optimal}
Nicolas Brunel, Vincent Hakim, Philippe Isope, Jean-Pierre Nadal, and Boris
  Barbour.
\newblock Optimal information storage and the distribution of synaptic weights:
  Perceptron versus purkinje cell.
\newblock {\em Neuron}, 43(5):745 -- 757, 2004.

\bibitem{Barbour_whatcanwelearn}
Boris Barbour, Nicolas Brunel, Vincent Hakim, and Jean-Pierre Nadal.
\newblock What can we learn from synaptic weight distributions?
\newblock {\em Trends in Neurosciences}, 30(12):622 -- 629, 2007.

\bibitem{Brunel_cortical}
Nicolas Brunel.
\newblock Is cortical connectivity optimized for storing information?
\newblock {\em Nature Neuroscience}, 19(5):749--755, 2016.

\bibitem{Gardner_spaceofinteractions}
E~Gardner.
\newblock The space of interactions in neural network models.
\newblock {\em Journal of Physics A: Mathematical and General}, 21(1):257--270,
  Jan 1988.

\bibitem{Clopath_storage}
Claudia Clopath, Jean-Pierre Nadal, and Nicolas Brunel.
\newblock Storage of correlated patterns in standard and bistable purkinje cell
  models.
\newblock {\em PLoS computational biology}, 8(4):e1002448--e1002448, 2012.

\bibitem{Chapeton_efficient}
Julio Chapeton, Tarec Fares, Darin LaSota, and Armen Stepanyants.
\newblock Efficient associative memory storage in cortical circuits of
  inhibitory and excitatory neurons.
\newblock {\em Proceedings of the National Academy of Sciences},
  109(51):E3614--E3622, 2012.

\bibitem{Zhang_associative}
Danke Zhang, Chi Zhang, and Armen Stepanyants.
\newblock Robust associative learning is sufficient to explain the structural
  and dynamical properties of local cortical circuits.
\newblock {\em Journal of Neuroscience}, 39(35):6888--6904, 2019.

\bibitem{RubinBalanced}
Ran Rubin, L.~F. Abbott, and Haim Sompolinsky.
\newblock Balanced excitation and inhibition are required for high-capacity,
  noise-robust neuronal selectivity.
\newblock {\em Proceedings of the National Academy of Sciences},
  114(44):E9366--E9375, 2017.

\bibitem{Seung_learning}
H.~S. Seung, H.~Sompolinsky, and N.~Tishby.
\newblock Statistical mechanics of learning from examples.
\newblock {\em Phys. Rev. A}, 45:6056--6091, Apr 1992.

\bibitem{Clopath_optimal}
Claudia Clopath and Nicolas Brunel.
\newblock Optimal properties of analog perceptrons with excitatory weights.
\newblock {\em PLOS Computational Biology}, 9(2):1--6, 02 2013.

\bibitem{Gutfreund_capacity}
H~Gutfreund and Y~Stein.
\newblock Capacity of neural networks with discrete synaptic couplings.
\newblock {\em Journal of Physics A: Mathematical and General},
  23(12):2613--2630, Jun 1990.

\bibitem{Isaacson_inhibition}
Jeffry~S. Isaacson and Massimo Scanziani.
\newblock How inhibition shapes cortical activity.
\newblock {\em Neuron}, 72(2):231 -- 243, 2011.

\bibitem{Field_HeterosynapticPlasticity}
Rachel~E. Field, James~A. D'amour, Robin Tremblay, Christoph Miehl, Bernardo
  Rudy, Julijana Gjorgjieva, and Robert~C. Froemke.
\newblock Heterosynaptic plasticity determines the set point for cortical
  excitatory-inhibitory balance.
\newblock {\em Neuron}, 2020.

\bibitem{Hennequin_review}
Guillaume Hennequin, Everton~J. Agnes, and Tim~P. Vogels.
\newblock Inhibitory plasticity: Balance, control, and codependence.
\newblock {\em Annual Review of Neuroscience}, 40(1):557--579, 2017.
\newblock PMID: 28598717.

\bibitem{ahmadian_dynamical}
Yashar Ahmadian and Kenneth~D. Miller.
\newblock What is the dynamical regime of cerebral cortex?
\newblock {\em arXiv:1908.10101}, 2019.

\bibitem{VanVreeswijkChaosScience}
C.~{van Vreeswijk} and H.~Sompolinsky.
\newblock Chaos in neuronal networks with balanced excitatory and inhibitory
  activity.
\newblock {\em Science}, 274(5293):1724--1726, 1996.

\bibitem{VanVreeswijkChaoticBalancedStateNeuralComputation}
C.~{van Vreeswijk} and H.~Sompolinsky.
\newblock Chaotic balanced state in a model of cortical circuits.
\newblock {\em Neural Comput.}, 10(6):1321--1371, Aug 1998.

\bibitem{RenartAsynchronous}
Alfonso Renart, Jaime de~la Rocha, Peter Bartho, Liad Hollender, N{\'e}stor
  Parga, Alex Reyes, and Kenneth~D. Harris.
\newblock The asynchronous state in cortical circuits.
\newblock {\em Science}, 327(5965):587--590, 2010.

\bibitem{KadmonSompolinsky}
Jonathan Kadmon and Haim Sompolinsky.
\newblock Transition to chaos in random neuronal networks.
\newblock {\em Phys. Rev. X}, 5:041030, Nov 2015.

\bibitem{HarishHanselAsynchronous}
Omri Harish and David Hansel.
\newblock Asynchronous rate chaos in spiking neuronal circuits.
\newblock {\em PLOS Computational Biology}, 11(7):1--38, 07 2015.

\bibitem{BrunelDynamicsSparsely}
Nicolas Brunel.
\newblock Dynamics of sparsely connected networks of excitatory and inhibitory
  spiking neurons.
\newblock {\em Journal of Computational Neuroscience}, 8(3):183--208, May 2000.

\bibitem{TsodyksStateSwitching}
M~V Tsodyks and T~Sejnowski.
\newblock Rapid state switching in balanced cortical network models.
\newblock {\em Network: Computation in Neural Systems}, 6(2):111--124, 1995.

\bibitem{goldt_modelling}
Sebastian Goldt, Marc M{\'e}zard, Florent Krzakala, and Lenka Zdeborov\'a.
\newblock Modelling the influence of data structure on learning in neural
  networks: the hidden manifold model.
\newblock {\em arXiv:1909.11500}, 2019.

\bibitem{Chung_perceptualmanifold}
SueYeon Chung, Daniel~D. Lee, and Haim Sompolinsky.
\newblock Classification and geometry of general perceptual manifolds.
\newblock {\em Phys. Rev. X}, 8:031003, Jul 2018.

\bibitem{Cohen_Separability}
Uri Cohen, SueYeon Chung, Daniel~D. Lee, and Haim Sompolinsky.
\newblock Separability and geometry of object manifolds in deep neural
  networks.
\newblock {\em Nature Communications}, 11(1):746, 2020.

\bibitem{Monasson_properties}
R~Monasson.
\newblock Properties of neural networks storing spatially correlated patterns.
\newblock {\em Journal of Physics A: Mathematical and General},
  25(13):3701--3720, Jul 1992.

\bibitem{Tarkowski_optimal}
Maciej Lewenstein and Wojciech Tarkowski.
\newblock Optimal storage of correlated patterns in neural-network memories.
\newblock {\em Phys. Rev. A}, 46:2139--2142, Aug 1992.

\bibitem{Monasson_correlatedpatterns}
R{\'e}mi {Monasson}.
\newblock {Storage of spatially correlated patterns in autoassociative
  memories}.
\newblock {\em Journal de Physique I}, 3(5):1141--1152, May 1993.

\bibitem{Battista_capacity}
Aldo Battista and R\'emi Monasson.
\newblock Capacity-resolution trade-off in the optimal learning of multiple
  low-dimensional manifolds by attractor neural networks.
\newblock {\em Phys. Rev. Lett.}, 124:048302, Jan 2020.

\bibitem{Gray_toeplitz}
Robert~M. Gray.
\newblock Toeplitz and circulant matrices: A review.
\newblock {\em Foundations and Trends in Communications and Information
  Theory}, 2(3):155--239, 2006.

\bibitem{abbott_interactions}
L~F Abbott, Kanaka Rajan, and Haim Sompolinsky.
\newblock Interactions between intrinsic and stimulus-evoked activity in
  recurrent neural networks.
\newblock {\em arXiv:0912.3832}, 2009.

\bibitem{LitwinKumar_optimal}
Ashok Litwin-Kumar, Kameron~Decker Harris, Richard Axel, Haim Sompolinsky, and
  L.F. Abbott.
\newblock Optimal degrees of synaptic connectivity.
\newblock {\em Neuron}, 93(5):1153 -- 1164.e7, 2017.

\bibitem{Marinari_replicafield}
E~Marinari, G~Parisi, and F~Ritort.
\newblock Replica field theory for deterministic models. {II}. a non-random
  spin glass with glassy behaviour.
\newblock {\em Journal of Physics A: Mathematical and General},
  27(23):7647--7668, Dec 1994.

\bibitem{Parisi_orthogonal}
G~Parisi and M~Potters.
\newblock Mean-field equations for spin models with orthogonal interaction
  matrices.
\newblock {\em Journal of Physics A: Mathematical and General},
  28(18):5267--5285, Sep 1995.

\bibitem{Cherrier_interactionmatrix}
R.~Cherrier, D.~S. Dean, and A.~Lef\`evre.
\newblock Role of the interaction matrix in mean-field spin glass models.
\newblock {\em Phys. Rev. E}, 67:046112, Apr 2003.

\bibitem{Opper_tractable}
Manfred Opper and Ole Winther.
\newblock Tractable approximations for probabilistic models: The adaptive
  thouless-anderson-palmer mean field approach.
\newblock {\em Phys. Rev. Lett.}, 86:3695--3699, Apr 2001.

\bibitem{Opper_adaptive}
Manfred Opper and Ole Winther.
\newblock Adaptive and self-averaging thouless-anderson-palmer mean-field
  theory for probabilistic modeling.
\newblock {\em Phys. Rev. E}, 64:056131, Oct 2001.

\bibitem{Opper_expectation}
M.~Opper and O.~Winther.
\newblock Expectation consistent approximate inference.
\newblock {\em Journal of Machine Learning Research}, 6:2177--2204, 2005.

\bibitem{Takeda_cdma}
K~Takeda, S~Uda, and Y~Kabashima.
\newblock Analysis of {CDMA} systems that are characterized by eigenvalue
  spectrum.
\newblock {\em Europhysics Letters ({EPL})}, 76(6):1193--1199, Dec 2006.

\bibitem{Kabashima_unified}
Y~Kabashima.
\newblock Inference from correlated patterns: a unified theory for perceptron
  learning and linear vector channels.
\newblock {\em Journal of Physics: Conference Series}, 95:012001, Jan 2008.

\bibitem{Shinzato_learning}
Takashi Shinzato and Yoshiyuki Kabashima.
\newblock Learning from correlated patterns by simple perceptrons.
\newblock {\em Journal of Physics A: Mathematical and Theoretical},
  42(1):015005, Nov 2008.

\bibitem{Shinzato_Revisited}
Takashi Shinzato and Yoshiyuki Kabashima.
\newblock Perceptron capacity revisited: classification ability for correlated
  patterns.
\newblock {\em Journal of Physics A: Mathematical and Theoretical},
  41(32):324013, Jul 2008.

\bibitem{Tulino_RandomMatrix}
Antonia~M. Tulino and Sergio Verd\'u.
\newblock Random matrix theory and wireless communications.
\newblock {\em Foundations and Trends in Communications and Information
  Theory}, 1(1):1--182, 2004.

\bibitem{Tao_topics}
T.~Tao.
\newblock {\em Topics in Random Matrix Theory}.
\newblock Graduate studies in mathematics. American Mathematical Soc., 2012.

\bibitem{Ganguli_compressed}
Surya Ganguli and Haim Sompolinsky.
\newblock Statistical mechanics of compressed sensing.
\newblock {\em Phys. Rev. Lett.}, 104:188701, May 2010.

\bibitem{Marr_cerebellar}
D.~Marr.
\newblock A theory of cerebellar cortex.
\newblock {\em The Journal of physiology}, 202(2):437--470, Jun 1969.

\bibitem{Wolpert_cerebellum}
Daniel~M Wolpert, R.Chris Miall, and Mitsuo Kawato.
\newblock Internal models in the cerebellum.
\newblock {\em Trends in Cognitive Sciences}, 2(9):338 -- 347, 1998.

\bibitem{Herzfeld_purkinje}
David~J. Herzfeld, Yoshiko Kojima, Robijanto Soetedjo, and Reza Shadmehr.
\newblock Encoding of error and learning to correct that error by the purkinje
  cells of the cerebellum.
\newblock {\em Nature Neuroscience}, 21(5):736--743, 2018.

\bibitem{mastrogiuseppeeinetworks}
Francesca Mastrogiuseppe and Srdjan Ostojic.
\newblock Intrinsically-generated fluctuating activity in excitatory-inhibitory
  networks.
\newblock {\em PLOS Computational Biology}, 13(4):1--40, 04 2017.

\bibitem{Chen_nonnegative}
J.~{Chen}, C.~{Richard}, J.~M. {Bermudez}, and P.~{Honeine}.
\newblock Variants of non-negative least-mean-square algorithm and convergence
  analysis.
\newblock {\em IEEE Transactions on Signal Processing}, 62(15):3990--4005, Aug
  2014.

\bibitem{Nascimento_rls}
V.~H. {Nascimento} and Y.~V. {Zakharov}.
\newblock Rls adaptive filter with inequality constraints.
\newblock {\em IEEE Signal Processing Letters}, 23(5):752--756, May 2016.

\bibitem{Engel_statistical}
Andreas Engel and Christian Van~den Broeck.
\newblock {\em Statistical mechanics of learning}.
\newblock Cambridge University Press, 2001.

\bibitem{mei_generalization}
Song Mei and Andrea Montanari.
\newblock The generalization error of random features regression: Precise
  asymptotics and double descent curve.
\newblock {\em arXiv:1908.05355}, 2019.

\bibitem{Gerace_generalisation}
Federica Gerace, Bruno Loureiro, Florent Krzakala, Marc M{\'e}zard, and Lenka
  Zdeborov\'a.
\newblock Generalisation error in learning with random features and the hidden
  manifold model.
\newblock {\em arXiv:2002.09339}, 2020.

\bibitem{Babadi_sparseness}
Baktash Babadi and Haim Sompolinsky.
\newblock Sparseness and expansion in sensory representations.
\newblock {\em Neuron}, 83(5):1213 -- 1226, 2014.

\bibitem{Cayco_patternseparation}
N.~Alex Cayco-Gajic and R.~Angus Silver.
\newblock Re-evaluating circuit mechanisms underlying pattern separation.
\newblock {\em Neuron}, 101(4):584 -- 602, 2019.

\bibitem{Ocker_motifs}
Gabriel~Koch Ocker, Ashok Litwin-Kumar, and Brent Doiron.
\newblock Self-organization of microcircuits in networks of spiking neurons
  with plastic synapses.
\newblock {\em PLOS Computational Biology}, 11(8):1--40, 08 2015.

\bibitem{Mezard_SpinGlassBeyond}
Marc M{\'e}zard, Giorgio Parisi, and Miguel Virasoro.
\newblock {\em Spin Glass Theory and Beyond}.
\newblock World Scientific Lecture Notes in Physics, 1987.

\end{thebibliography}

\end{document}